\documentclass[11pt,journal,onecolumn,draftclsnofoot]{IEEEtran}
\usepackage{amsmath,amsfonts}
\usepackage[linesnumbered,ruled,vlined]{algorithm2e} 
\usepackage{lineno} % To obtain line numbers
\usepackage{array}
\usepackage{subcaption}
\usepackage{textcomp}
\usepackage{stfloats}
\usepackage{url}
\usepackage{verbatim}
\usepackage{graphicx}
\usepackage{cite}
\def\fn{\footnote} 
\usepackage{xcolor}
\usepackage{mathtools}
\mathtoolsset{showonlyrefs}
\usepackage{multirow}
\usepackage{makecell}
\usepackage{geometry}
\usepackage{float}

\usepackage{booktabs} % For pretty tables

\DeclareRobustCommand*{\IEEEauthorrefmark}[1]{
\raisebox{0pt}[0pt][0pt]{\textsuperscript{\footnotesize\ensuremath{#1}}}}

\usepackage[colorlinks=true,linkcolor=blue]{hyperref}
\usepackage{xr}
\begin{document}
% \linenumbers

\title{Glider Path Design and Control for \\ 
Reconstructing Three-Dimensional Structures of \\ 
Oceanic Mesoscale Eddies}

% \author{Wu Su, Xiaoyuan E, Zhao Jing, Song Xi Chen}

\author{
\IEEEauthorblockN{
Wu Su\IEEEauthorrefmark{1},
Xiaoyuan E\IEEEauthorrefmark{2},
Zhao Jing\IEEEauthorrefmark{3, 4, *}, and
Song Xi Chen\IEEEauthorrefmark{5, 6, *}} \\
\vspace{3mm}
\IEEEauthorblockA{\IEEEauthorrefmark{1}Center for Big Data Research, Peking University, Beijing, China}\\
\IEEEauthorblockA{\IEEEauthorrefmark{2}JD Logistics, Beijing, China}\\
\IEEEauthorblockA{\IEEEauthorrefmark{3}Frontiers Science Center for Deep Ocean Multispheres and Earth System and Key Laboratory of Physical Oceanography, Ocean University of China, Qingdao, China.}\\
\IEEEauthorblockA{\IEEEauthorrefmark{4}Laoshan Laboratory, Qingdao, China.}\\
\IEEEauthorblockA{\IEEEauthorrefmark{5}Center for Statistical Science, Peking University, Beijing, China}\\
\IEEEauthorblockA{\IEEEauthorrefmark{6}School of Mathematical Sciences and Guanghua School of Management, Peking University, Beijing, China}\\
\IEEEauthorblockA{\IEEEauthorrefmark{*}{\bf Corresponding Authors}: Zhao Jing, jingzhao@ouc.edu.cn; Song Xi Chen, csx@gsm.pku.edu.cn}}



\maketitle

\begin{abstract}

Underwater gliders offer effective means in oceanic surveys with a major task in reconstructing the three-dimensional hydrographic field  of a mesoscale eddy. This paper considers three key issues in the hydrographic  reconstruction of  mesoscale eddies with the sampled data from the underwater gliders. It  first proposes using the Thin Plate Spline (TPS) as the interpolation method for the reconstruction with a blocking scheme to speed up the computation.  It then formulates a procedure for selecting  glider path design that  minimizes the reconstruction errors among a set of pathway formations. Finally we provide  a glider path control procedure to guide the glider to follow to designed pathways as much as possible in the presence of ocean current. A set of optimization algorithms are experimented and several with robust glider control performance on a simulated eddy are identified.    

\end{abstract}


\section{Introduction} \label{sec: Intro}

The mesoscale eddy is one of the most dominant forms of motion in the ocean with diameter ranging from tens to hundreds of kilometers and a vertical depth of hundreds to thousands 
meters \cite{CHELTON2011167}. The study of the  
three-dimensional hydrographic field of mesoscale eddies is a major issue in  oceanography and marine  science with great significance in the study of climate and marine ecosystems  \cite{RN22, Waterman2011, Gaube2014, doi:10.1126/science.1252418, ImpactsonOceanHeatfromTransientMesoscaleEddiesinaHierarchyofClimateModels, RN23, Rohr2020s, https://doi.org/10.1029/2019GB006385, doi:10.1126/sciadv.aba7880}.
 
Although the spatial resolution of state-of-the-art ocean general circulation models has become fine enough to resolve mesoscale eddies due to the rapidly increased computational capacity  \cite{gmd-13-4595-2020}, they are still far from being able to accurately simulate 
mesoscale eddies \cite{RN17, MORETON2020101567}. Simulating a particular mesoscale eddy is even more challenging 
in practice due to its turbulent properties. The remote-sensing based measurements can only  
acquire the footprint of mesoscale eddies at the sea surface, such as eddy-induced sea surface height and temperature anomaly  \cite{5898391}. 
The surface eddy signals can be used to infer the three-dimensional structure of mesoscale eddies  \cite{RN18, RN19, RN20, RN21}. For instance, \cite{https://doi.org/10.1002/grl.50736} argued that the vertical structure of mesoscale eddies can be universally represented by a sinusoidal function in a stretched coordinate depending on the ratio of buoyancy frequency to Coriolis frequency. \cite{https://doi.org/10.1029/2007JC004692}  reconstructed the three-dimensional structure of mesoscale eddies using the sea surface temperature from satellite observations based on a surface quasi-geostrophic framework. However, the dynamical frameworks underpinning the reconstruction are likely to be an oversimplification of the real ocean. Furthermore, the reconstruction is essentially done for seawater density. The temperature and salinity associated with mesoscale eddies cannot be derived without making assumptions on their relationship. 

Controllable and autonomous ocean in-situ observation platforms provide an alternative solution to observe the eddy structure by actively collecting the hydrographic data within mesoscale eddies, and then reconstructing the three-dimensional hydrographic field of mesoscale eddies from the collected data. An underwater glider (UG) is an autonomous underwater vehicle (AUV), which 
adjusts its buoyancy to realize rising and glides 
with the help of hydrodynamics of ocean current using fixed wings  \cite{2018Development, Yang2019, Wang2024}. 

The glider collects data such as sea water temperature  and salinity with quite high temporal resolution along designed paths. By recording the underwater speed of the sensor, the glider's three-dimensional coordinates can be calculated \cite{ma2019absolute}. Due to its relatively low cost and large spatial coverage, underwater gliders have been widely used to observe the three-dimensional hydrographic structure of mesoscale eddies, and the need for high-resolution {reconstruction of mesoscale eddies} has seen more gliders  being  put into use.  
 
{There are three  imperative issues in efficient usage of ocean gliders. One is  how to design the travel paths of the gliders in order to provide efficient sampled data for more accurate reconstruction on the three-dimensional hydrographic field of a mesoscale eddy. 
The other one is how to make  the glider to travel along the designed pathway in the presence of ocean current.  The third issue is in  finding  computationally efficient interpolation methods for efficient accomplishing  the three-dimensional hydrographic properties reconstruction based on the glider sampled data.  There has been no a systematic solution to these issues in existing research as far as we are aware. } 
 
{For the glider path design, the existing works tended to focus on  minimizing  two-dimensional reconstruction errors. For the circular glider formations, \cite{2007Collective} derived the glider path scheme with minimal expected reconstruction error at two-dimensional sea surface. \cite{4209085} combined gliders with a network of profiling floats to calculate the optimal path design for up to three gliders using a genetic algorithm to minimize the average error of the reconstructed field. A combination of mooring systems was likewise considered in \cite{Alvarez2012}. } 

A popular method for the reconstruction of mesoscale eddy is the spatial interpolation  based on smooth two-dimensional or three-dimensional interpolation functions  \cite{https://doi.org/10.1029/2007JC004692,2016Observed}.  
 \cite{10.1145/800186.810616} proposed the inverse distance weighted (IDW) interpolation, which  
assigns larger weights to the closer observed data points. Another commonly used spatial interpolation method is the Kriging \cite{MatheronKrig1963, 1992Precipitation, NALDER1998211, Cressiech3}, which is a statistical method that utilizes the spatial dependence for best linear unbiased  interpolation. The IDW and Kriging methods are much used in geology and atmospheric science, and the IDW interpolation had been used in  three-dimensional  eddy reconstruction  \cite{LI2020101893}.

Interpolation with spline functions is another method, which fits a piece-wise function through observation points and satisfy certain smoothness conditions  \cite{HutchinsonSmooth}. \cite{SomeNew} introduced spline interpolations to spatial analysis in meteorology, and in particular  the Thin Plate  Spline (TPS). \cite{HUTCHINSON199445} proved that there is a mathematical connection between the TPS  and the Kriging, and pointed out that the TPS interpolation using generalized cross-validation to select the roughness penalty parameters may be more efficient than Kriging. 
\cite{ThinPlateSmoothingSplineModelingofSpatialClimateDataandItsApplicationtoMappingSouthPacificRainfalls} proposed a method to select tuning parameters by minimizing  the mean square error, which further improved the performance of the TPS interpolation.

Existing  studies on glider path control were largely focused on getting the glider to the destination with the  least amount of  time and energy consumption, without paying much attention on the glider's path design and the quality of the eddy reconstruction.   
As the glider's movement is interfered by the ocean current, control methods such as the Proportional–Integral–Derivative (PID) controller have been used in attitude control of gliders.  \cite{2018Heading} proposed a hybrid heading control algorithm that integrated an  adaptive fuzzy incremental PID with an anti-saturation compensation strategy to achieve robust heading control.  Other studies considered minimizing  the travel time or energy consumption  \cite{Garau2009PathPF}. \cite{BESADAPORTAS2013111} applied the particle swarm optimization algorithm to glider path controlling. \cite{2014Differential} conducted a series of simulation experiments and field tests. The feasibility of applying Differential Evolution (DE) Algorithm to the task of minimizing the travel time of the glider was made in  \cite{2016Constrained, Zamuda2017Adaptive, 2019Success}. 
In addition, the reinforcement learning method has been considered. It divides 
the spatial locations into grid points, which inevitably leads to a loss of resolution; the use of this method in high resolution situations remains to be investigated \cite{8604754RL1, 9705882RL2}. 
{Furthermore, these approaches require information of the current field and they actually did not perform glider control on an optimally designated path.} 

This paper addresses three major issues with the use of gliders for oceanic data collection with the purpose of 
reconstructing the hydrographic field of a 
mesoscale eddy. We will demonstrate that  the TPS method is an effective interpolation method and propose a three dimensional blocking scheme to speed up the computation of the TPS interpolation based on the glider sampled data.  Secondly, we propose a data-driven glider path design 
%of the underwater gliders 
to find the best formation pattern so that using the sampled data along the formation would achieve the smallest reconstruction error among a set of candidate pathway  formations. A center and a parallel path designs are showed to be advantageous for a simulated eddy.  
After having designed the glider travel  paths, we consider glider control to ensure that they travel along the designed path as much as possible in the presence of ocean current. We  devise an objective function that is a weighted sum of the glider's absolute deviation to the designated pathway and the distance to the final destination with the weights being adjustable according to the magnitude of the ocean current. Having the distance to the destination in the objective function is to prevent the glider being stuck or lost all together under  strong ocean current, which can happen if the objective function is only based on the deviation to the designed path.  Eight optimization algorithms including the DE and Self-adaptive DE (JDE) methods are experimented on the simulated eddy with the observed ocean current, and robust performing path control algorithms are identified.  

The rest of the paper is organized as follows. Section \ref{Glider interpolation} introduces TPS interpolation and proposes a three-dimensional blocking interpolation method for the rapid reconstruction of the hydrographic field of an eddy. Sections \ref{Glider Path Design} and \ref{Glider Path Control} present the proposed glider path design and path control methods, respectively. Experimental results on the simulated eddy are presented in Sections \ref{Experimental Results} and \ref{Control Experimental Results}, followed by a conclusion in Section \ref{Conclusions}. Additional technical and numerical details are given in the supplementary material (SM).

\section{Interpolation for Hydrographic Reconstruction} 
\label{Glider interpolation} 

\begin{figure}[h]
    \centering
    \includegraphics[width = \textwidth]{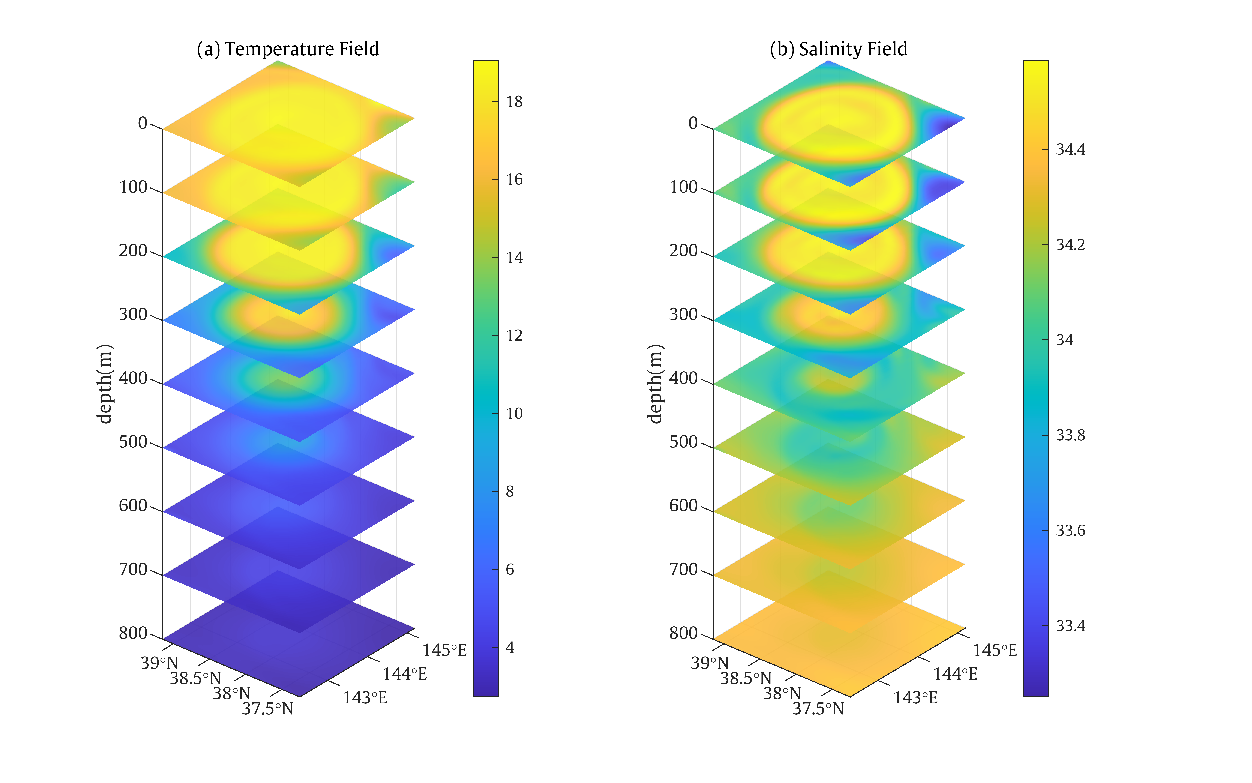}
    \caption{Three dimensional structure of the temperature ($^\circ \text{C}$) and salinity (g/kg) fields of a simulated mesoscale eddy, where the three axes represent longitude, latitude and depth, respectively.}
    \label{fig: TSField}
\end{figure}

This section presents the TPS based method for reconstructing the  hydrographic field using sampled data from the gliders, and proposes a three-dimensional blocking algorithm to speed up the TPS computation. {Fig \ref{fig: TSField} displays the hydrographic structure of a simulated eddy, which shows  strong three-dimensional spatial dependence.} 

\subsection{Thin Plate  Spline}

Wahba \& Wendelberger introduced the TPS interpolation to spatial data analysis in meteorology \cite{SomeNew}.  \cite{https://doi.org/10.1002/joc.1276} employed the TPS to interpolate meteorological variables over global land area to achieve higher spatial resolution at the time.

For sampled data points $\left( \boldsymbol{X}_i,Y_i \right), i = 1,2,\cdots, n$, where  $\boldsymbol{X}_i$ is the input variable of $p$ dimensions and $Y_i$ is the response variable. { In our context, $\boldsymbol{X}_i$ is three-dimensional coordinate consisting of longitude, latitude, and depth, while $Y_i$ represents the temperature or salinity at $\boldsymbol{X}_i$.} We aim at finding the best interpolation function $f^*(x)$ at $\mathbf{x} =\left( x^{\left( 1 \right)},\cdots ,x^{\left( p \right)} \right) ^T$ that minimizes the sum of squared interpolation errors along with the bending energy {which quantifies the smoothness of a surface} \cite{24792}, namely 
\begin{equation} \label{eq: TPS, J obj} 
 \mathcal{J}_n\left( f \right) = \sum_{i=1}^n{\left( Y_i-f\left( \boldsymbol{X}_i \right) \right) ^2}  + \lambda \int_{\mathcal{R}^p}{\left( \sum_{i=1}^p{\sum_{j=1}^p{\left( \frac{\partial ^2f}{\partial x^{\left( i \right)}x^{\left( j \right)}} \right) ^2}} \right) \text{d}x^{\left( 1 \right)}\cdots \text{d}x^{\left( p \right)}}. 
\end{equation} 
The integral above is the two order bending energy of $f$, and $\lambda > 0$ is a smoothing parameter. A larger (smaller) $\lambda$ will result in a smoother (rougher) interpolation function. 

This leads to the optimization problem
%\begin{equation} \label{eq: TPS, objective function}
$\min_f\  {\mathcal{J}}_n\left( f \right)$.   
%\end{equation}
\iffalse
The  function $f$ that minimizes $\mathcal{J}_n$ must satisfy the
Euler-Lagrange equation $D\mathcal{J}_n\left( f \right) =0$, where, as shown in the SM,   
\begin{equation} \label{eq: TPS, first variation}
    D\mathcal{J}_n\left( f \right) =2\sum_{i=1}^n{\left( Y_i-f\left( \boldsymbol{X}_i \right) \right) \delta \left( \mathbf{x} -\boldsymbol{X}_i \right)}+2\lambda \Delta ^2f\left( \mathbf{x}  \right), 
\end{equation}
where $\delta \left( \cdot \right)$ is the Dirac delta function and 
\begin{equation}
    \Delta ^2f=\sum_{i=1}^p{\sum_{j=1}^p{\left( \frac{\partial ^4f}{\partial \left( x^{\left( i \right)} \right) ^2\partial \left( x^{\left( j \right)} \right) ^2} \right)}}.
\end{equation}

The first order condition for the optimization \eqref{eq: TPS, objective function} is 
    \begin{equation} \label{eq: TPS, E-L}
\sum_{i=1}^n{\left( f\left( \boldsymbol{X}_i \right) -Y_i \right) \delta \left(\mathbf{x}  -\boldsymbol{X}_i \right)}+\lambda \Delta ^2f(\mathbf{x} )=0,  
    \end{equation}
which can be solved via the Green functions (\cite{ Duchon1976Interpolation, 4767807})  in the form of  
        \begin{equation} \label{eq: Green}
            G_p\left( r \right) =\begin{cases}
	r^{4-p}\ln r,&		p=2,4;\\
	r^{4-p},&		\text{otherwise}.\\ 
\end{cases}
        \end{equation}
\fi 
As shown in Section \ref{supp-sec:TPS} of SM, the  minimizer $f^*$  is  
\begin{equation} \label{eq: TPS, interpolation function}
    f^*\left( \mathbf{x}  \right) =\beta _0+\sum_{k=1}^p{\beta _kx^{(k)}}+\sum_{i=1}^n{w_iG_p\left( \lVert \mathbf{x}-\boldsymbol{X}_i \rVert _2 \right)}, \, \hbox{where} \quad G_p\left( r \right) =\begin{cases}
	r^{4-p}\ln r,&		p=2,4;\\
	r^{4-p},&		\text{otherwise}.\\ 
\end{cases}, 
\end{equation}        
the coefficients $\beta=\left(\beta _0,\beta _1,\cdots ,\beta _p\right)^T$ and the weights $w=(w_1,\cdots,w_n)^T$ are  to be estimated. Derivations given in the SM  show that $\boldsymbol{\beta}$ and $\mathbf{w}$ can be obtained by solving 
\begin{equation} \label{eq: TPS, linear system}
\left[ \begin{matrix}
	\mathbf{A}+\lambda I,&		\mathbf{X}\\
	\mathbf{X}^T,&		O\\
\end{matrix} \right] \left( \begin{array}{c}
	\mathbf{w}\\
	\boldsymbol{\beta}\\
\end{array} \right) =\left( \begin{array}{c}
	\mathbf{Y}\\
	\mathbf{0}\\
\end{array} \right), 
    \end{equation}
  where $\mathbf{Y}=\left( Y_1,\cdots ,Y_n \right) ^T$, the $i$-th row of matrix $\mathbf{X}$ is $\left[ 1, \boldsymbol{X}_i^T \right]$, for $i=1,2,\cdots n$, and $\mathbf{A} = (a_{ij})$ with the entries $a_{ij}=G_p\left( \lVert \boldsymbol{X}_i-\boldsymbol{X}_j \rVert _2 \right)$ for $i \ne j$ and $a_{ii}=0$.  
    
Let $\hat{\boldsymbol{\beta}}$ and $\hat{\mathbf{w}}$ be the estimates  after solving equation \eqref{eq: TPS, linear system}. Suppose one wants to interpolate at a coordinate  $\boldsymbol{X}_0 = \left( X_{0,1},X_{0,2},X_{0,3} \right)$.  The interpolation  by the TPS is % can be calculated by the following formula
\begin{equation} \label{eq: TPS, 3d}
\hat{Y}_0=f^*\left( \boldsymbol{X}_0 \right) =\hat{\beta}_0+\hat{\beta}_1X_{0,1}+\hat{\beta}_2X_{0,2}+\hat{\beta}_3X_{0,3}+\sum_{i=1}^n{\hat{w}_i r_{0 i} }, 
\end{equation}
where $r_{0 i}$ is the Euclidean distance between $\boldsymbol{X}_0$ and $\boldsymbol{X}_i$. 

It is noted that despite we engage a three dimension reconstruction, what is found is that using the two-dimensional Green function $G_2(r)=r^2 \ln r$ brings better empirical results than those using the three dimensional function, which had been found in other studies (\cite{jmse8110832, 8835020}). This may be due to the fact that although the purpose is for the three-dimensional reconstruction, the design is largely two dimensional via line segments across the sea surface. Hence, we use the two dimensional Green function in our study.

\subsection{Blocking Strategy for Efficient Computation} \label{subsec: block strategy}

It is noted that the time complexity for solving the linear system \eqref{eq: TPS, linear system} is $\mathcal{O}\left(n^3 \right)$ since $p\ll n$. As the data sample collected by gliders is usually more than 100,000, the TPS  interpolation will endure heavy computation burden. As pointed out by \cite{ThinPlateSmoothingSplineModelingofSpatialClimateDataandItsApplicationtoMappingSouthPacificRainfalls}, the number of observations for the conventional TPS interpolation is better to be less than 1000 to ensure  computation. Hence, it is necessary to develop a  computation  efficient  algorithm for timely TPS interpolation for the three-dimensional eddy's temperature or salinity field.  

We propose a blocking strategy, which divides the entire three dimensional reconstruction space into  overlapping cuboids.  
For convenience, the study area $R$ is assumed to be a unit cube with unit edge length,  which can  be realized by the min-max scaling, namely $x_{i}^{*}=\frac{x_i-\min \left\{ x_i \right\}}{\max \left\{ x_i \right\} -\min \left\{ x_i \right\}}$, so that  $R=\left\{ \left( x_1,x_2,x_3 \right) \mid 0\le x_1,x_2,x_3\le 1 \right\}$ where $x_1,x_2$ and $x_3$ represent the standardized longitude, latitude and depth. We divide the unit interval $[0,1]$ corresponding to each dimension to form multiple  sub-cuboids. Denote the number of blocks in longitude, latitude and depth are 
$B_{\text{long}},B_{\text{lat}}$ and $B_{\text{dep}}$, respectively. Take longitude as example, the $i$-th sub-interval with an overlapping ratio $c$ is  $\left[ E_{\text{long,}i}^{L},E_{\text{long,}i}^{U} \right], i = 1,2,\cdots, B_{\text{long}}$, where 
\begin{equation}\label{eq: sub-interval}
\left\{ \begin{array}{l}
	E_{\text{long,}i}^{L}=\max \left\{ 0,\frac{i-1 - \frac{c}{2}}{B_{\text{long}}} \right\} ,\,\,\\
	E_{\text{long,}i}^{U}=\min \left\{ 1,\frac{i + \frac{c}{2}}{B_{\text{long}}} \right\} ,\\
\end{array} \right. 
\end{equation}
while $\left[ E_{\text{lat,}i}^{L},E_{\text{lat,}i}^{U} \right]$ and $\left[ E_{\text{dep,}i}^{L},E_{\text{dep,}i}^{U} \right]$ can be defined similarly. Then, the cube $R$ is divided to $B=B_{\text{long}}\cdot B_{\text{lat}}\cdot B_{\text{dep}}$ sub-cuboids, denoted as 
\begin{equation}\label{eq: sub-cuboids} 
R_{ijk}=\left[ E_{\text{long,}i}^{L},E_{\text{long,}i}^{U} \right] \times \left[ E_{\text{lat,}j}^{L},E_{\text{lat,}j}^{U} \right] \times \left[ E_{\text{dep,}k}^{L},E_{\text{dep,}k}^{U} \right] ,
\end{equation}
for $i=1,\cdots ,B_{\text{long}};j=1,\cdots ,B_{\text{lat}};k=1,\cdots ,B_{\text{dep}}$. 

We first build the TPS interpolation function $\hat{f}_{ijk}\left( \mathbf{x} \right)$ using  all the sampled observations in the sub-cuboid $R_{ijk}$. The final interpolation function is a weighted average of $\hat{f}_{ijk}(\mathbf{x})$ over all the sub-cuboids $\{R_{ijk}\}$: 
\begin{equation}\label{eq: weighted interpolation function}
    \hat{f}\left( \mathbf{x} \right) =\sum_{i=1}^{B_{\text{long}}}{\sum_{j=1}^{B_{\text{lat}}}{\sum_{k=1}^{B_{\text{dep}}}{\omega _{ijk}\left( \mathbf{x} \right) \hat{f}_{ijk}\left( \mathbf{x} \right)}}}, 
\end{equation}
where the weights $\omega_{ijk}(\mathbf{x})$ are determined as follows.  

\begin{figure}[h]
            \centering
            \includegraphics[width= 0.47 \textwidth]{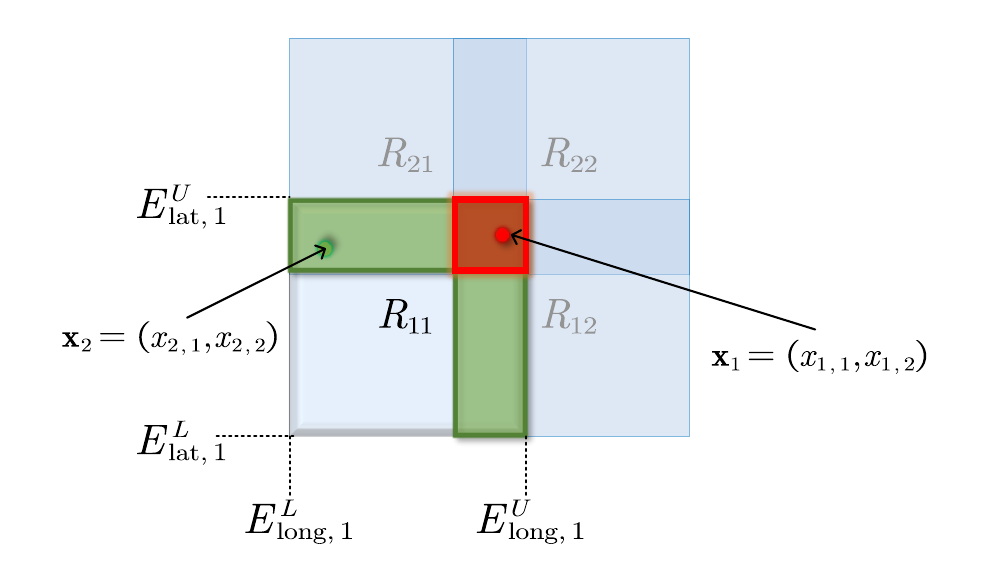}
            \caption{Two dimensional demonstration of the blocking scheme. For $R_{11}$, the green parts are the overlaps of two regions and the red part is the overlap of four regions. 
            $d_{11}\left( \mathbf{x}_1 \right) = | x_{1,1}-E_{\text{long,}1}^{U}  | | x_{1,2}-E_{\text{lat,}1}^{U}  |$,  $d_{11}\left( \mathbf{x}_2 \right) = | x_{2,1}-E_{\text{long,}1}^{L}  | | x_{2,2}-E_{\text{lat,}1}^{U}  |$ and $d_{12}\left( \mathbf{x}_2 \right) = d_{22}\left( \mathbf{x}_2 \right) = 0$. 
            }
            \label{fig: block stategy}
\end{figure}

Let  $\mathbf{x}=\left( x_1,x_2,x_3 \right)$ be the location where the interpolation is made. For all $R_{ijk}$ that contain $\mathbf{x}$,  define
\begin{equation}\label{eq: sij}
\left\{ \begin{array}{l}
	s_{\text{long,}ijk}\left( \mathbf{x} \right) =\min \left\{ | x_1-E_{\text{long,}i}^{L} |,| x_1-E_{\text{long,}i}^{U} | \right\} ,\\
	s_{\text{lat,}ijk}\left( \mathbf{x} \right) =\min \left\{ | x_2-E_{\text{lat,}j}^{L}|,| x_2-E_{\text{lat,}j}^{U} | \right\} ,\\
	s_{\text{dep,}ijk}\left( \mathbf{x} \right) =\min \left\{| x_3-E_{\text{dep,}k}^{L} |,| x_3-E_{\text{dep,}k}^{U} | \right\},\\
\end{array} \right.  
\end{equation}
which measure the shortest distance from $\mathbf{x}$ to the boundaries of $R_{ijk}$. Furthermore, let 
\begin{equation} \label{eq: di} 
d_{ijk}\left( \mathbf{x} \right) =\begin{cases}
	s_{\text{long,}ijk}\left( \mathbf{x} \right) \cdot s_{\text{lat,}ijk}\left( \mathbf{x} \right) \cdot s_{\text{dep,}ijk}\left( \mathbf{x} \right) ,&		\mathbf{x}\in R_{ijk},\\
	0,&		\mathbf{x}\notin R_{ijk}. \\
\end{cases} 
\end{equation} 
Fig \ref{fig: block stategy} shows a two-dimensional blocking and how $d(\mathbf{x})$ is calculated. 

After obtaining $d_{ijk}(\mathbf{x})$, the weight 
\begin{equation}\label{eq: wi}
\omega _{ijk}\left( \mathbf{x} \right) =\frac{d_{ijk}^{2}\left( \mathbf{x} \right)}{\sum_{u=1}^{B_{\text{long}}}{\sum_{v=1}^{B_{\text{lat}}}{\sum_{w=1}^{B_{\text{dep}}}{d_{uvw}^{2}\left( \mathbf{x} \right)}}}}. 
\end{equation}
It can be seen that when $\mathbf{x} \in R_{ijk}$, the closer (more away) $\mathbf{x}$ is to the boundary of $R_{ijk}$, the smaller (larger) the weight $\omega_{ijk}(\mathbf{x})$. When $\mathbf{x}$ does not fall within $R_{ijk}$, $\omega_{ijk}(\mathbf{x})$ is 0. Thus, the final interpolated value at $\mathbf{x}$ depends only on those interpolation functions corresponding to those sub-cuboids  that contain $\mathbf{x}$. The idea of making the sub-cuboids overlapping is to maintain smoothness around the boundaries. Furthermore, squaring $d_{ijk}(\mathbf{x})$  in the weight $\omega_{ijk}(\mathbf{x})$ also helps to maintain smoothness of the interpolation function.

The blocking strategy effectively reduces computation time of the interpolation method.  The computational complexity of the full TPS depends on solving the linear system \eqref{eq: TPS, linear system}, which is $\mathcal{O}(n^3)$. If there are roughly $n/B$ observations in each block,  the computational complexity of the blocking scheme is $B(n/B)^3 = B^{-2} n^3$, implying a factor of $1/B^2$ less computation relative to the full sample approach.

Algorithm \ref{supp-alg:alg1} in SM provides  the three-dimensional blocking  interpolation algorithm. It is noted that the blocking algorithm can  be applied to other spatial interpolation methods for instance the IDW and the Kriging.

\section{Glider Path Design} \label{Glider Path Design}

\begin{figure}[p]
  \centering
  \includegraphics[width= 0.8 \textwidth]{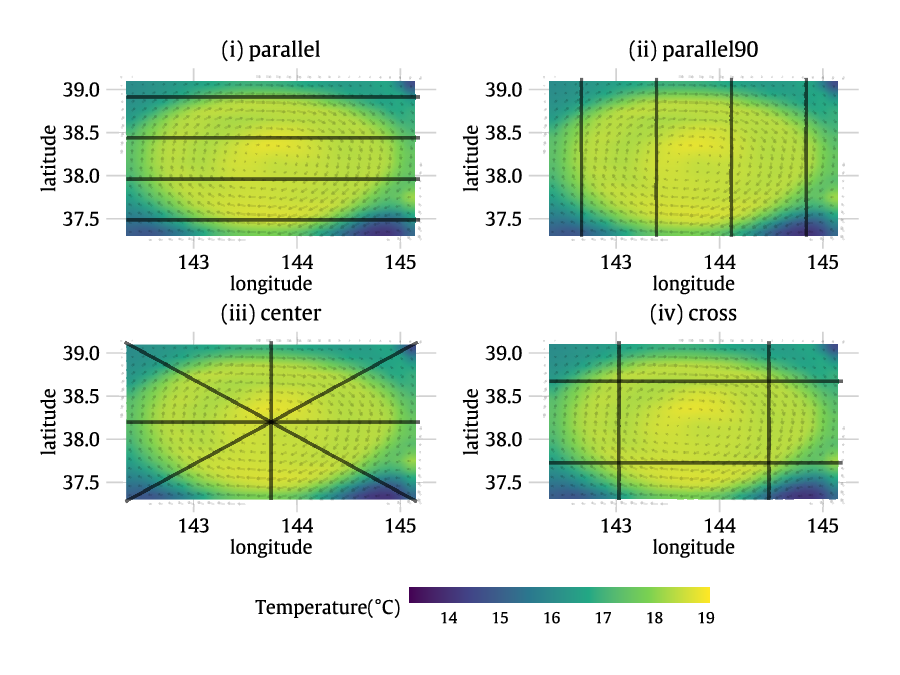} \vskip -30pt
  \subfloat[Different glider pathways  for 4 gliders \label{fig: topological structures}]{\hspace{\linewidth}}
  
  \vspace{3mm}
  
  \includegraphics[width=  0.8 \textwidth]{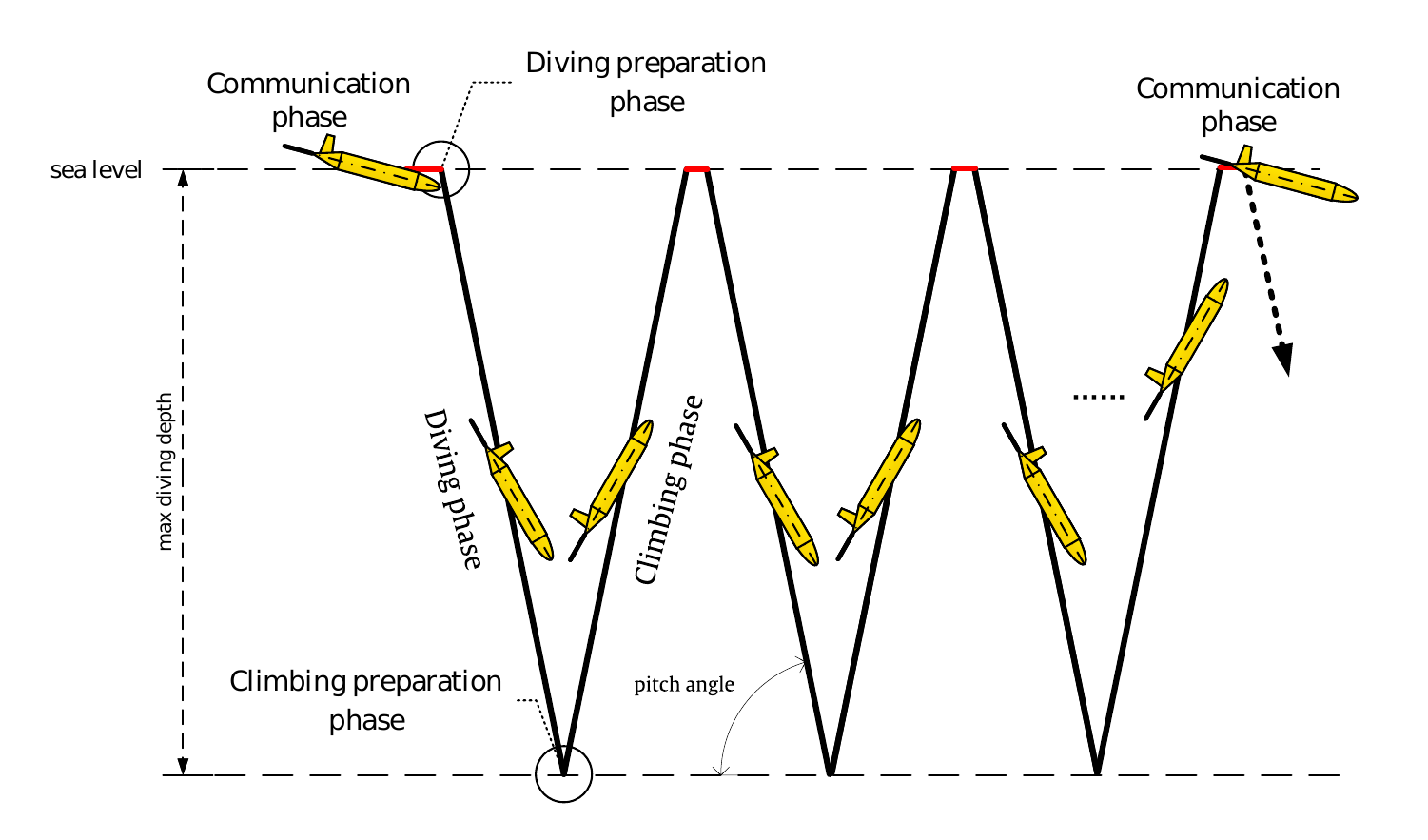} \vskip -25pt
  \subfloat[ Repeated diving and climbing motions of the underwater glider \label{fig:diving and climbing}]{\hspace{\linewidth}}

  \caption{(a) The projection of the underwater gliders' trajectory on the sea level (4 different path designs) and (b) its diving and climbing motions on the vertical section. \label{fig: above and below}}
\end{figure}

{We consider how to efficiently allocate the glider travel paths within the study region so that the sampled data from the gliders lead to more accurate reconstruction of the hydrographic field of the eddy. } 
In this study, the glider paths are ``line segments" \cite{Yang2023, Wang2024} when viewed vertically down from the sky 
with repeated diving to a specified depth and climbing to the sea surface following straight line segments as well, see  Fig \ref{fig: above and below} for illustration. This is consistent with the commonly adopted path patterns for  gliders  \cite{2018Coordinate, LI2020101893}. 
As the purpose of the glider path design is for the reconstruction of the three-dimensional hydrographic field of the mesoscale eddy, the goal is to find a design pattern for the line segments that 
minimizes the reconstruction error based on data collected along the designated paths. 

Suppose there are  $K$  gliders % , and there are 
and a set of glider path formations denoted as $\{\mathcal{A}_j\}_{j=1}^M$  at our disposal. Fig \ref{fig: above and below} presents four examples of  path formations for $K=4$ gliders: 
\begin{itemize}
    \item parallel: $K$ parallel lines   passing through the eddy in the longitude direction;
    \item parallel90: $K$ parallel lines passing  through the eddy in the meridian direction; 
    \item center: $K$ lines all passing through the estimated eddy center with equal rotation angles;
    \item cross:  two sets (each having $K/2$ gliders)  of parallel lines, which are  mutually perpendicular.
\end{itemize}
The above cross formation is for even number of $K$ and can be extended to odd $K$ with near half splitting of the horizontal and vertical sets. Considering that it is impractical for gliders to carry out non-linear paths, one may consider only linear path formations and find the optimal solution within easy to implement path schemes in field trials as shown in Fig \ref{fig: topological structures}. 

The purpose of the path design is to find the best path formation with the minimum reconstruction error among all candidate formations $\{\mathcal{A}_j\}_{j=1}^M$  based on a training data set.  
The training data set can be obtained from satellite remote sensing measurements {of an eddy’s sea surface temperature (SST) or salinity (SSS), using sources such as OSTIA for SST \cite{OSTIA} and SMOS for SSS \cite{SMOS}, } which may be augmented by a high-resolution ocean  analysis {product} (e.g. \url{https://www.hycom.org/}) that provides estimates of three-dimensional hydrographic field by combining the numerical model and observations via data assimilation  \cite{Cummings2005, Cummings2013}. 
%\fn{maybe we can mention aviso etc. namely update this paragraph (Wu: AVISO provides SSH, but here we talk about temp/salt. Maybe we can mention OSTIA for SST and SMOS for SSS. done)}  
Although these estimates are subject to estimation error, 
they provide useful information on the overall structure of the temperature and salinity filed of the eddy and can be used to guide the glider path design. In the empirical study reported later, we use a training data generated from a high-resolution community earth system model (CESM; \cite{CESM2014}) with an oceanic resolution of 10km.
%\fn{I thought you have a total of 5 eddies. May mention the effects of the MEs (Wu: I add 2 sentences at the end of this paragraph)} 
%Although there is no one-to-one correspondence between the CESM simulated eddies  and the observed ones,\fn{what is this for ? (Wu: Prof JZ write this part. This means that there is no direct correspondence between the simulated eddies and those observed in reality. I then went on to mention that we considered reanalysis datasets, which have a certain degree of correspondence with real-world eddies.)} 
Using the CESM data is suitable for this study as it allows establishing a universal framework for the optimal eddy sampling path design by gliders. {To demonstrate the method's applicability to real-world situation, in addition to the simulated eddy, we incorporated three more active eddies in the Kuroshio Extension region from August 12, 2024, together with the GLORYS reanalysis data for the hydrographic field over the period. These real-world-matched eddies allow us to test the proposed approach’s effectiveness in practical implementation and its robustness. } 
%in the presence of measurement errors.}

It is worth mentioning that the data collection by the gliders  is not only determined by the sampling path, but also by factors such as glider speed, maximum diving depth, pitch angle and sampling interval. We regard these later factors to be given in the path design.

For a glider path formation $\mathcal{A}_j$, the $K$ gliders collect information such as the temperature or salinity $Y_i$ at locations $\boldsymbol{X}_i = (X_{i,1},X_{i,2},X_{i,3})$  for $i = 1,2,\cdots,n$ along the line paths. 
Then we use the sampled  data $\{(X_i, Y_i)\}_{i=1}^n$ to fit an interpolation function $\hat{f}_j^*(\mathbf{x})$, 
and evaluate the interpolation error on a test data set $\mathcal{T}=\left\{ \left( \boldsymbol{Z}_1,T_1 \right) ,\cdots ,\left( \boldsymbol{Z}_{N},T_{N} \right) \right\}$. The reconstruction Root Mean Square Error (RMSE) is given by
\begin{equation}\label{eq: rec error}
    \text{RMSE}({\mathcal{A}_j}) = \sqrt{\frac{1}{N}\sum_{t = 1}^{N}{\left( \hat{f}^*_j\left( \boldsymbol{Z}_t \right) -T_t \right) ^2}}. 
\end{equation}

Glider path design is to find a design $\mathcal{A}^*$ among the  candidate set $\{\mathcal{A}_j\}_{j=1}^M$, which minimizes the reconstruction RMSE among all  formations designs $\{\mathcal{A}_j\}_{j=1}^M$, namely, 
\begin{equation}\label{eq: min error} 
\mathcal{A}^*=\operatorname*{argmin}_{j = 1,2,\cdots, m}  \text{RMSE}({\mathcal{A}_j}).  
\end{equation}

\section{Methods for Glider Path Control} \label{Glider Path Control} 

In this section, we present an adaptive path control algorithm to make the gliders follow the designed  pathways in the presence ocean current. The path control of an underwater glider faces the challenge of  knowing its real-time position only after its climbing to sea surface by the GPS system and the interference of ocean current.  The path control is conducted by adjusting the heading angle of the next diving and climbing, based on the current position, the ocean current information and the designed line paths. 

{The Proportional–Integral–Derivative (PID) controller is a popular method used for dynamic system control. A PID controller continuously calculates the difference $e(t)$ between the designated and the true positions of the object to be controlled after each time $t$ and applies a correction based on the  proportional (P), integral (I) and derivative (D) of $e(t)$. The overall control function is 
\begin{equation} \label{eq: PID}
    u\left( t \right) =K_Pe\left( t \right) +K_I\int_0^t{e\left( s \right) \mathrm{d}s}+K_D\frac{\mathrm{d}e\left( t \right)}{\mathrm{d}t},
\end{equation}
where $K_P,K_I,K_D$ are non-negative coefficients. In the case of gliders, $u(t)$ will be used to correct for the heading angle.  
}  \cite{2018Heading} proposed an adaptive fuzzy incremental PID control to dynamically adjust the heading angles; see also \cite{Liu2017} for a similar approach. 
As the PID algorithm is weak against disturbances, it may not work well under strong ocean current \cite{Ullah2015}. 

In addition to the PID method, there were studies using  optimization methods for glider path control \cite{BESADAPORTAS2013111,2014Differential}. However, as mentioned in Section \ref{sec: Intro}, these studies formulated the glider path control problem as a shortest path  (in fact the shortest time) problem,  and in addition it  requested that the ocean current field was known over the entire control period.

For the task of mesoscale eddy reconstruction, we want the gliders to travel along the designed paths unless the ocean current is too strong to make it impossible.  We present an optimization framework for the glider path control based on the ocean current information  by adjusting the heading angle after each glider surfacing.  

Considering that we can interact with the glider only when it  surfaces, and the glider path design is largely two dimensional in terms of  sea surface plane, we may represent the path control  as a  two-dimensional problem ignoring the depth. Suppose the starting and destination positions of a designed line path are, respectively,  $\boldsymbol{x}_{\text{start}}=\left( x_{\text{start,}1},x_{\text{start,}2}\right)$ and  $\boldsymbol{x}_{\text{target}}=\left( x_{\text{target,}1},x_{\text{target,}2} \right)$ on the sea surface. We assume that the parameters such as glider's speed, the pitch angle and the maximum dive depth are fixed for a whole mission. 
%Therefore, the control variable is the heading angle, or we consider the turning angle $\alpha$ at each climbing up to the surface waiting for the instruction of the next move. 
{Therefore, the control variable is the heading angle. Specifically, each time the glider surfaces and waits for instructions for its next move, we determine the turning angle $\alpha$ to adjust its direction for the subsequent path segment.}
 
Let $\boldsymbol{x}_i$, $i = 0,1,2,\cdots$, be the position on the sea surface of glider after $i$ diving and climbing motions. At each $\boldsymbol{x}_i$, we want to decide the next turning angle $\alpha_{i + 1}$. Then,  one has to  consider what control strategy is optimal for the next $H$ diving and climbing motions, rather than just for one step ahead. The next $H$ positions of glider $\{\boldsymbol{x}_{i+1}, \cdots, \boldsymbol{x}_{i+H}\}$ depend on the turning angles $\alpha_{i+1},\cdots,\alpha_{i+H}$.  

To find the best turning angles, we design the objective function as follows.  Firstly, we want the realized glider path to follow the  designed pathway with the method proposed in Section \ref{Glider Path Design}. A measure of the fidelity of the realized  and the designed paths is  the average deviation between the $H$ surfacing positions $\{\boldsymbol{x}_{i+l}\}_{l=1}^H$ and the designed path line denoted as $\ell :Ax_1+Bx_2+C=0$, {where $A$, $B$, and $C$ are the coefficients defining the straight line connecting the starting position $\boldsymbol{x}_{\text{start}}$ and the destination position $\boldsymbol{x}_{\text{target}}$ in the plane.} This leads to the first part of the objective function: 
\begin{equation} \label{eq: deviation} 
   f_1\left( \alpha _{i+1},\cdots ,\alpha _{i+H} \right) =\frac{1}{H}\sum_{j={i+1}}^{i+H}{\frac{\left| Ax_{j,1}+Bx_{j,2}+C \right|}{\sqrt{A^2+B^2}}}. 
\end{equation}
While minimizing $f_1\left( \alpha _{i+1},\cdots ,\alpha _{i+H} \right)$ would ensure the travel pathway is close to the designed one, however, under strong ocean current, the glider may stuck at a place or being blow away from the survey area. This suggests we have to find a compromise between the fidelity to the designed pathway and the overall objective to reach the final destination. To this end, we construct the second  objective function 
\begin{equation}\label{eq:minD1_rewrite}
f_2\left( \alpha _{i+1},\cdots ,\alpha _{i+H} \right)  =\lVert \boldsymbol{x}_{\text{target}}-\boldsymbol{x}_{i+H} \rVert_2 .
\end{equation}
Minimizing this objective function means having the $H$ ahead position of the glider as close to the destination point $\boldsymbol{x}_{\text{target}}$. The overall objective function is a weighted sum of the two: 
\begin{equation} \label{eq:weighted average}
    f\left(  \alpha _{i+1},\cdots ,\alpha _{i+H} \right) =  w_1f_1\left( \alpha _{i+1},\cdots ,\alpha _{i+H}\right)+ w_2f_2\left(  \alpha _{i+1},\cdots ,\alpha _{i+H} \right), \\
\end{equation}
where the weights $w_1$ and $w_2$ depend  on the ocean current condition. Generally speaking, if the ocean current  is  weak (strong), we should make $w_1$ larger (smaller). 
Detailed instructions will be given shortly.

\begin{figure}[!htb]
    \centering
    \includegraphics[width = 0.47 \textwidth]{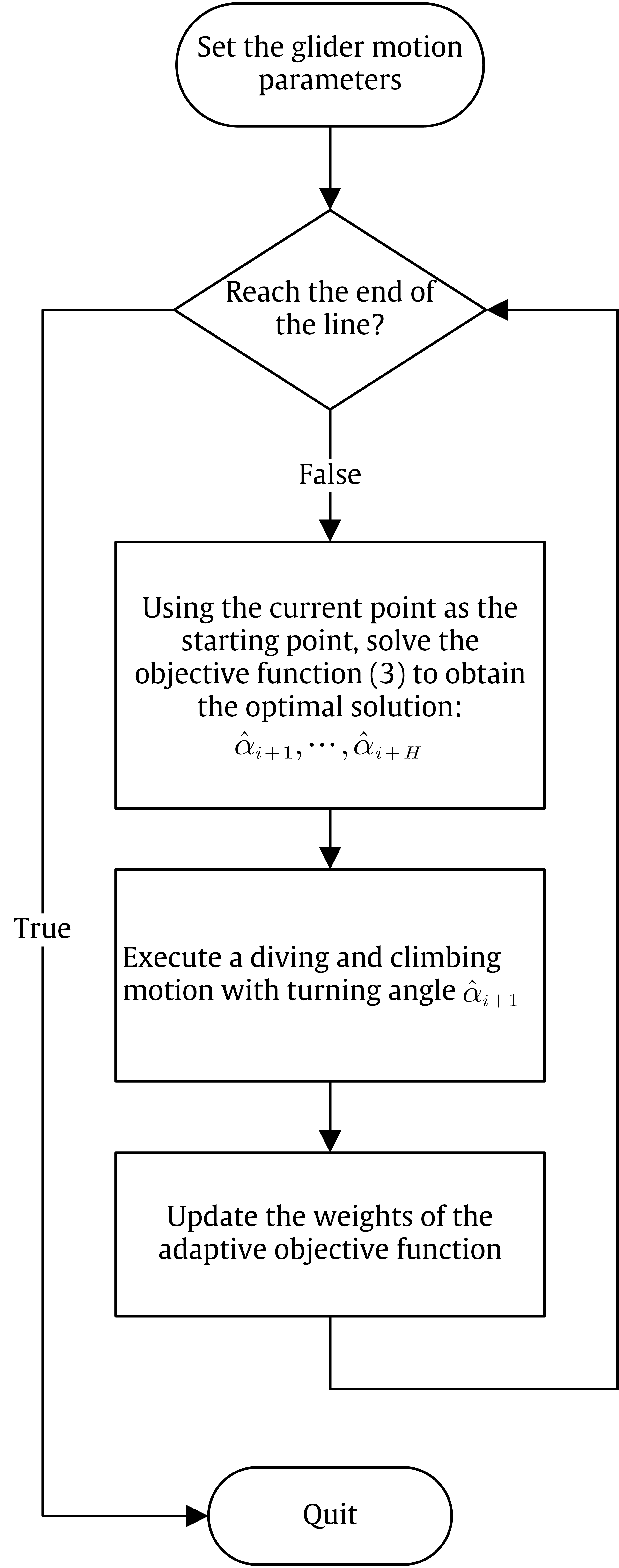} \vskip -10pt
    \caption{Flowchart of glider path control, where the optimization problem  solves the adaptive objective function \eqref{eq:weighted average}. }
    \label{fig: flowchart}
\end{figure}

It is noted that although the result from solving \eqref{eq:weighted average} at $\boldsymbol{x}_i$ is $(\hat{\alpha}_{i+1},\cdots, \hat{\alpha}_{i+H})$, we only use $\hat{\alpha}_{i+1}$ to correct the current heading angle. And in the next step,  we  repeat the $H$-step ahead optimization again but based at position $\boldsymbol{x}_{i+1}$. 
  The reason for our not directly using one solution of optimizing  \eqref{eq:weighted average}  for the complete path 
  is that that would require a very large $H$ being  used and  the complexity of the optimization would  be very large. It is easy to optimize with a relatively small $H$. We chose $H=10$ in the implementation. 
  Also a large $H$ would demand more on the accuracy of the ocean current field over a longer time horizon.  Fig \ref{fig: flowchart} displays the algorithmic framework for  the main steps in the glider path control problem, where the optimization of \eqref{eq:weighted average} is called  the glider path control sub-problem. 

Minimizing \eqref{eq:weighted average} is a derivative-free optimization problem, which means that a heuristic algorithm like the Differential Evolution (DE) method (see \ref{supp-sec:de} of SM  for details) has to be applied. In order to reduce the search space, we limit the turning angle to a range   $\left[ -100^{\circ},100^{\circ} \right]$ in the implementation and fix  $H = 10$. The value of $H$ can change with respect to the ocean current condition  or if the  glider is near its destination,  the end of the line path. 

The weights in the objective function \eqref{eq:weighted average} are updated each time  when the glider surfaces. Let $\boldsymbol{x}_{k-1}$ and $\boldsymbol{x}_k$ be the positions where the glider surfaces for the $(k-1)$-th and $k$-th times, let
\begin{equation} \label{eq: relaDis}
r_{k-1}=\frac{\lVert \boldsymbol{x}_{k-1}-\boldsymbol{x}_{\text{target}} \rVert}{\lVert \boldsymbol{x}_{\text{start}}-\boldsymbol{x}_{\text{target}} \rVert} \quad \hbox{and} \quad r_k=\frac{\lVert \boldsymbol{x}_k-\boldsymbol{x}_{\text{target}} \rVert}{\lVert \boldsymbol{x}_{\text{start}}-\boldsymbol{x}_{\text{target}} \rVert}
\end{equation}
be the relative distance of the two locations to the destination, respectively. Then let
\begin{equation}\label{eq:Deltar}
    \Delta r_k=\max \left( r_{k-1}-r_k,\varepsilon \right),  
\end{equation}
where $r_{k-1}-r_k$ is the relative distance reduction  to the destination  and $\varepsilon $ 
is a small positive number, for instance  $10^{-5}$, to prevent $\Delta r_k$ being negative.  If  $\Delta r_k$ 
is  close to $\varepsilon $, it indicates that the glider is hardly moving forward or even retreating, and we should decrease $w_1$ and increase $w_2$. On the contrary, if $\Delta r_k$ is large, it implies that the glider goes smoothly, and $w_2$ can be decreased while $w_1$ can be increased to make the glider sample along the target path. 

Next, we discuss how to update $w_1$ and $w_2$.  First define $c_k= \frac{\delta}{\Delta r_k}$, 
where   
$\delta$ is a tuning parameter that reflects  the average relative travel distance per diving cycle.
{The expected number of diving and climbing cycles under static current field is $\frac{L\tan \beta}{2h}$, where $L$ is the length of the designed glider path, $h$ is the diving depth and $\beta$ is the pitch angle. These  mean  that the relative distance traveled  per diving cycle is $\frac{2h}{L\tan \beta}$. 
However, considering the presence of ocean current, one may choose $\delta =\frac{h}{L\tan \beta}$. 
}
Now the weights are updated according to 
\begin{equation}\label{eq:wk update}
    w_{1,k+1}= c_{k}^{-1}w_{1,k}, \quad
w_{2,k+1}=c_k w_{2,k} . 
\end{equation}
We would bound the weight updating factor $c_k$ to prevent too drastic weight changes. 
%in the weights from erratic movement of the glider.  
For that purpose, let
\begin{equation} \label{eq: ckstar}
    c_{k}^{*}=\begin{cases}
	c_{\min},&	\hbox{if} \,	c_k\le c_{\min},\\
	c_k,&	  	\hbox{if} \, c_{\min}<c_k\le c_{\max},\\
	c_{\max},&	\hbox{if} \,	c_k>c_{\max}.\\
\end{cases}
\end{equation}
One may set 
$c_{\min}$ and $c_{\max}$ to be the reciprocal of each other, for instance $c_{\min} = 0.1$ and $c_{\max} = 10$.  
We also introduce $w_{1,\min}, w_{1,\max}$ and $w_{2,\min}, w_{2,\max}$ to restrain $w_1$ and $w_2$ {from both ends}, respectively, to prevent one of the objective functions from being too dominant. {Without such bounds, the weights could become excessively large or small under certain ocean conditions.} %\fn{R4Q9} 
Algorithm \ref{alg:alg2} outlines the algorithm for the adaptive glider path control. 

\begin{algorithm}[ht]
\caption{Adaptive glider path control\label{alg:alg2}}
\footnotesize
\LinesNumbered 
\KwIn{ $\boldsymbol{x}_{\text{start}}, \boldsymbol{x}_{\text{target}}$, glider initial position $\boldsymbol{x}_0$, the dimension of search space $H_0$, endpoint tolerance $\eta$,  and several parameters in adaptive adjustment: $c_{\min},c_{\max},w_{1,\min},w_{1,\max},w_{2,\min},w_{2,\max},\varepsilon ,\delta$. }
Initialize the heading angle $\phi_0$ toward $\boldsymbol{x}_{\text{target}}$ \;
Initialize weights: $$ 
w_{1,1}\gets \frac{w_{1,\min}+w_{1,\max}}{2},w_{2,1} \gets \frac{w_{2,\min}+w_{2,\max}}{2};$$ 

Initialize search dimension $ H \gets H_0$ \;
$k \gets 1$ \;
\While{the glider is still in the study area or has not exceeded its maximum sailing time}{
    Given $\left( \boldsymbol{x}_{\text{start}},\boldsymbol{x}_{\text{target}} \right)$, and the position of glider $\boldsymbol{x}_{k - 1}$, use heuristic algorithm (e.g. DE) to solve the following optimization problem: 
$$
\left( \hat{\alpha}_{k},\cdots ,\hat{\alpha}_{k+H-1} \right)^T \gets \mathop{\arg\min}\limits_{ -100^{\circ} \le \alpha _i \le 100^{\circ}}w_{1,k}f_1 +w_{2,k}f_2 .
$$

    Update the heading angle $\phi_{k} \gets \phi_{k-1} + \hat{\alpha}_{k}$ and use it for the next diving and climbing motion \;
    Gets the current surface position $\boldsymbol{x}_{k}$ of the glider \;
    \If{ $\lVert \boldsymbol{x}_k-\boldsymbol{x}_{\text{target}} \rVert \le \eta $}{
        \textbf{Break} \;
    }
    Calculate $r_{k-1} \gets \frac{\lVert \boldsymbol{x}_{k-1}-\boldsymbol{x}_{\text{target}} \rVert}{\lVert \boldsymbol{x}_{\text{start}}-\boldsymbol{x}_{\text{target}} \rVert},r_k \gets \frac{\lVert \boldsymbol{x}_k-\boldsymbol{x}_{\text{target}} \rVert}{\lVert \boldsymbol{x}_{\text{start}}-\boldsymbol{x}_{\text{target}} \rVert}$ \;
    $\Delta r_k  \gets \max \left( r_{k-1}-r_k,\varepsilon \right)$ \;
    $ c_k \gets \max \left\{ c_{\min},\min \left\{ \frac{\delta}{\Delta r_k},c_{\max} \right\} \right\}$ \;
    Update weights:
    $$
    \begin{aligned}
    w_{1,k+1}& \gets \max \left\{ w_{1,\min},\min \left\{ c_{k}^{-1}w_{1,k},w_{1,\max} \right\} \right\} ,
\\
w_{2,k+1}&\gets \max \left\{ w_{2,\min},\min \left\{ c_kw_{2,k},w_{2,\max} \right\} \right\} .
    \end{aligned}
    $$

\If{$w_{2,k+1} = w_{2,\max}$}{
    $H \gets H + H_0$\;
}\Else{
    $H \gets \max\{H_0, H - H_0\}$ \;
}
    $H \gets \min \left\{ H,\lceil \frac{r_k}{\Delta r_k} \rceil \right\}$ \;
    $k \gets k + 1$ \;
}
\end{algorithm}  

Considering that the velocity field of the ocean current is only available for the surface, one has to estimate of the current field beneath the surface, which can be made via the ocean models or the historical data by establishing  the relationship between the ocean surface current and the ocean interior current. This aspect will be discussed in Section \ref{subsec: EVF}.

\section{Experimental Results for Path Design} \label{Experimental Results}

This section reports results from numerical evaluation of the proposed procedure based on a simulated mesoscale eddy from CESM on March 18, 1989 over a region  with latitude 37.25N-39.15N and longitude 142.3E-145.2E. The spatial resolution of the simulated mesoscale eddy  was $0.1^\circ \times 0.1^\circ$, and there were 39 levels from the sea surface to 830m below, which makes $30\times20\times39=23400$ grid boxes in the three-dimension for the eddy. 

The simulated eddy was regarded as the truth and reconstruction accuracy of its temperature and salinity field was the focus of the evaluation.  %active, strong eddies which are subject to measurement errors.}  
{The goal of the path design is to determine the number of gliders and their travel paths %according to the 
%providing guidance for the actual mission. }
%Specifically, we  
by evaluating the numerical performance of the glider path designs proposed in Section \ref{Glider Path Design} based on the simulated eddy without interference of ocean current.  } 
{This may be regarded as a design based on the  hydrographic field of an eddy at a time, and the design can be updated upon given an updated hydrographic 
%\fn{does it contain the current ? (Wu: No need here, since we consider path design in ideal calm ocean)}  
field. 
This  requires fast computation in selecting the glider's travel path, which is the rationale for employing the TPS with the blocking approach. In practice, the hydrographic field may be obtained from a real-time numerical model or data assimilation system \cite{Cummings2005, Oke2008}, like CESM or GLORYS. } 
%\fn{mention CESM ? (Wu: OK, but CESM is just a numerical model rather than DA system. I %mention both CESM and GLORYS)}
%\fn{mention aviso ? (Wu: AVISO provide SSH, but we do not need SSH here. I have mentioned it in the path control %phase.)}
{In additional to %Although our discussion is based on this  
the simulated eddy, we demonstrate  on the  three eddies in the Kuroshio extension
%\fn{So they are different from the three from Kuroshio extension ? (Wu: The same three) %}  
from GLORYS  that  our approach can be applied to time-evolving eddies as shown in Section \ref{supp-sec:res-design} of SM.} %  \fn{tell (Wu: OK, done)}.}
%{\bf 
% Although the eddy evolves over time, we focus on an effective design when a hydrographic field that reflects the eddy structure is available. This hydrographic field may be obtained from a real-time data assimilation system \cite{Cummings2005, Oke2008}. The design can be immediately updated whenever an updated hydrographic field is available. This necessitates rapid computation for selecting the glider's travel path, which is the rationale for employing the TPS with the blocking approach.
%} %\fn{some repeat, may reduce a bit  (Wu:OK, I have shorten it)}
%\fn{R1Q2, R4Q10. time variation; R1Q2, model partially reflects reality, using Data Assimilation}
% Numerical evaluations in the presence of the ocean current will be the focus of Section \ref{Control Experimental Results}. 
 
For the simulation of the glider sampling, we refer to the Petrel-II %developed by China\fn{too vague. Can you provide a ref ?},  
for simulation settings. Specifically, we set the optional range of pitch angle as $20^\circ$, the data sampling interval was once every 50 seconds(s), the horizontal velocity component at 0.5m/s, the vertical velocity component 0.2m/s   \cite{2010Coordinated,Scripps2015Ocean}, and  the maximum diving depth  to be 800m,  which was also determined by the  ocean depth in the research area.   

\subsection{Comparing Different Interpolation Methods} \label{sec:interpolation}

Before evaluating the effectiveness of the proposed glider path design and control, we first compare the performance of three interpolation methods including the inverse distance weighting (IDW), the Kriging  and the TPS interpolations. In addition, we tried two blocking strategies. One had three-dimensional overlapping blocks (3D) as described in Algorithm \ref{supp-alg:alg1},  where we set $B_{\text{long}} = 3, B_{\text{lat}} = 2$ and  $B_{\text{dep}} = 4$ with an overlap ratio between blocks $c = 0.25$; and the other had the blocks confined in  the two-dimensional plane at a given depth (2D), namely the two-dimensional interpolation was performed on each of the 39 depths.  For the IDW interpolation method, we set the power parameter $p=5$ for both the block and plane versions according to the cross-validation results. The smoothing parameter $\lambda$ used by the TPS was automatically estimated by minimizing the generalized cross-validation (GCV) score \cite{Wahba_book, ThinPlateSmoothingSplineModelingofSpatialClimateDataandItsApplicationtoMappingSouthPacificRainfalls}. Kriging interpolation 
%\fn{say a few words on its dependence on a covariance function ? wait: the covariance can be estimated}  
was implemented using MATLAB's {\sf fitgpr} function, with the default Gaussian kernel selected. The parameters of the kernel function were determined using the built-in cross-validation method.

\begin{table}[ht]
      \caption{The root mean square errors (RMSEs) of the reconstructed  eddy hydrographic field using three interpolation methods:  the inverse distance weighted (IDW), the Kriging and the proposed Thin Plate Spline (TPS) with  the two blocking schemes:  the two dimensional (2D) and three (3D) dimensional blocking schemes.} 
      \label{tab: Interpolation Comparison}
        \centering
\begin{tabular}{lcc}
\toprule
\multirow{2}{*}[-.6mm]{Method} & \multicolumn{2}{c}{\textbf{$\text{RMSE}_{\text{CV}}$}} \\ \cmidrule(lr){2-3} 
                        & Temperature($^\circ \mathrm{C}$)  & Salinity(g/kg)  \\ \midrule
IDW.2D               & 0.2648465    & 0.0533172 \\
IDW.3D               & 0.2153072    & 0.0450347 \\
Kriging.2D           & 0.1305336    & 0.0261869 \\
Kriging.3D           & 0.1132758    & 0.0237654 \\
TPS.2D               & 0.3065255    & 0.0641885 \\
TPS.3D               & \textbf{0.1027353}    & \textbf{0.0203141} \\ \bottomrule
\end{tabular}
\end{table}

Table \ref{tab: Interpolation Comparison} reports the root mean square errors (RMSEs) from the 5-fold cross-validation where the $30\times20\times39=23400$ grids eddy data were randomly partitioned to five portions 
{with four portions used to train the interpolation model, while  calculating the prediction RMSEs on the remaining one. This cross validation  operation was repeated five times to obtain the cross-validated RMSEs} for the three interpolation methods coupled with the two  blocking schemes. 
In interpolation with  the two dimensional blocking (2D) method, 
the kriging  attained the smallest RMSEs, followed by the IDW and TPS.  
The reason for the poor performance of the TPS with plane blocking was due to that the temperature and the salinity field of the eddy have strong vertical dependence as shown in Fig \ref {fig: TSField}, and the TPS interpolation tends to give smoother  estimates than the other two methods  while the slices of plane blocks remove much of the vertical dependence. 
However, when the vertical dependence was included in the 3D blocking scheme,  the TPS interpolation exhibited  its advantage over the other two methods.  
Indeed, the block TPS interpolation attained  the smallest RMSEs in both the temperature and salinity reconstruction. 
The 3D TPS  had  9.3\% and 14.5\%  less RMSEs for   temperature and salinity  reconstruction, respectively,  than those of  the Kriging method and much more than those of the IDW. 

{% The above analysis is based on the uniform sampling of data. Additionally, 
To gain insights on the the performance of the interpolation methods  under sparse glider sampling and the presence of measurement errors, we consider the three  eddies on August 12, 2024 in the Kuroshio Extension region. 
%which are obtained from GLORYS reanalysis data.
%We also considered three more active eddies from August 12, 2024, in the Kuroshio Extension region, obtained from GLORYS reanalysis data.  
Using data sampled in the glider paths, we reconstructed the  hydrographic fields of these eddies and assessed the impact of measurement errors from glider sampling. In this setting, the training data were unevenly distributed and noisier. 
\iffalse 
\fn{what does this mean ? (Wu: uniform sampling means we randomly split grid points to training/testing data. And glider sparse sampling means we generate glider sampling data with parallel/parallel90 design then reconstruct the whole  hydrographic field. The latter one implies the training distribution is uneven and allows the inclusion of MEs. ) SX:  then, the idea is not properly written. (Wu: I have modified  it)} 

\fn{which three eddies ? need some information on them (Wu: OK, I have added some information)} 
\fi 
Figure \ref{supp-fig:three_eddies} displays the spatial range, sea surface height and surface current of the three eddies. Figures \ref{supp-fig:eddy1_res}-\ref{supp-fig:eddy3_res} present  the reconstruction error and runtime of each interpolation algorithm in reconstructing the eddies. 
%\fn{say something on what are these figures present (Wu: OK, done)} 
It is evident that the Kriging interpolation, due to its reliance on covariance structure assumption, 
%\fn{true ? (Wu: I think it maybe this case. Since it is challenging to learn and generalize covariance from uneven distribution, though it may not be straightforward to derive this conclusion theoretically)}  
performed inconsistently under the sparse sampling situation. In contrast, the TPS interpolation had more stable performance for each eddy under the same sparse and noisy data, while having a significant advantage in the computational efficiency. A more detailed discussion is provided in Section \ref{supp-sec:interp} of the SM. } %\fn{R1Q1, discussion about TPS and Kriging. R1Q2, measurement error}

In the rest of the simulation, only the 3D TPS interpolation was considered.

\subsection{Path Design} \label{Path Design}

We compare reconstruction performance with different path designs and different number of gliders. The path design with four gliders is shown in Fig \ref{fig: topological structures}.  Four to ten  gliders were used in the simulation experiments using the design procedure described  in Section \ref{Glider Path Design}.

Section \ref{subsec: block strategy} has discussed the computation time with respect to  the number of blocks. It can be seen from Fig \ref{fig: topological structures} that the glider's sample coverage in the latitude and longitude plane was quite sparse with quite some remote areas from the allocated glider paths. 
Hence,  the number of blocks with respect to  the longitude and latitude should not be too large, and  we set $B_{\text{long}} = B_{\text{lat}} = 3$. {Given that the glider's CTD sensor can sample at high frequency, there is effectively no extra sparsity with respect to depth.} %\fn{R4Q8} 
Therefore, we choose the number of blocks in the vertical direction ranging from 10, 20, 40, to 80 in the simulation experiments.

Table \ref{tab:zong} summarizes the reconstruction RMSEs  of the eddy temperature and salinity fields, the glider sampling effort in travel length and {the correlation between the values obtained from the TPS interpolation and the actual values of the underlying eddy field,} %\fn{R3Q2}, 
with respect to different number of gliders. It also presents results with respect to the number of vertical layer blocks $B_\text{dep}$ and different path design patterns.   

Table \ref{tab:zong} (a) and (b)  show that  as the number of gliders increased, the reconstruction RMSEs almost monotonically reduced for each given pattern of path designs  for both temperature and salinity field reconstruction. 
The parallel and the center path designs offered the least RMSEs among the four patterns for temperature, and the center design had the least RMSE for salinity.  
{Notably, the center design required the longest travel distance, whereas the parallel90 design required the least. Despite its greater survey effort, the center design did not always achieve the lowest RMSEs for the temperature field, nor did the parallel90 design consistently yield the lowest RMSEs. The cross design appeared less effective than the parallel90, as it had higher RMSEs despite a longer travel distance.}
Additionally note that the reconstruction under the center design with 4 gliders encountered a singular coefficient matrix in \eqref{eq: TPS, linear system},  which led to no result for the case. %\fn{(Wu: I have rewritten for clarify)}

\begin{table}[p] \footnotesize
   \centering

   \caption{(a) Temperature ($^\circ \mathrm{C}$) field and (b) Salinity (g/kg) field reconstruction RMSE and correlation coefficient (Corr.) with their respective true values for different path designs and different number of gliders along with the length of the designated paths (in km); (c) Temperature ($^\circ \mathrm{C}$) field reconstruction root mean square error (RMSE) and CPU time (in second or minute or hour) with  different number of blocks. }

   \vskip -3mm
   \subfloat[Temperature field reconstruction performance with $B_{\text{long}} = B_{\text{lat}} = 3$ and $B_{\text{dep}} = 40$\label{tab: Temp reconstruction}]{
\begin{tabular}{ccccccccccccc}
\toprule
\multirow{3}{*}[-2mm]{  \makecell{number \\ of \\ gliders}} & \multicolumn{12}{c}{Glider Path Design}                                                                                 \\ \cmidrule[.5pt]{2-13} 
                                     & \multicolumn{3}{c}{parallel} & \multicolumn{3}{c}{parallel90} & \multicolumn{3}{c}{center} & \multicolumn{3}{c}{cross} \\
                                     \cmidrule(rl){2-4} \cmidrule(rl){5-7} \cmidrule(rl){8-10} \cmidrule(rl){11-13}
                                     & length   & RMSE    & Corr.   & length   & RMSE     & Corr.    & length  & RMSE    & Corr.  & length  & RMSE   & Corr.  \\ \midrule[.5pt]
4  & 1027 & 0.3831          & 0.9979          & 845  & \textbf{0.3643} & \textbf{0.9981} & 1449 & /               & /               & 936  & 0.6130 & 0.9945 \\
5  & 1283 & \textbf{0.2435} & \textbf{0.9991} & 1056 & 0.2786          & 0.9989          & 1449 & 0.2890          & 0.9988          & 1193 & 0.4224 & 0.9974 \\
6  & 1540 & 0.1960          & 0.9994          & 1268 & 0.2252          & 0.9992          & 1872 & \textbf{0.1881} & \textbf{0.9995} & 1404 & 0.3125 & 0.9986 \\
7  & 1797 & \textbf{0.1343} & \textbf{0.9997} & 1479 & 0.1636          & 0.9996          & 1872 & 0.1667          & 0.9996          & 1661 & 0.2091 & 0.9994 \\
8  & 2053 & 0.1124          & 0.9998          & 1690 & 0.1319          & 0.9997          & 2385 & \textbf{0.1005} & \textbf{0.9999} & 1872 & 0.1578 & 0.9996 \\
9  & 2310 & 0.0946          & \textbf{0.9999} & 1901 & 0.1124          & 0.9998          & 2385 & \textbf{0.0911} & \textbf{0.9999} & 2128 & 0.1177 & 0.9998 \\
10 & 2567 & \textbf{0.0716} & \textbf{0.9999} & 2113 & 0.0884          & \textbf{0.9999} & 2808 & 0.0822          & \textbf{0.9999} & 2340 & 0.0943 & \textbf{0.9999} \\ \bottomrule
\end{tabular}}

   \subfloat[Salinity field reconstruction performance with $B_{\text{long}} = B_{\text{lat}} = 3$ and $B_{\text{dep}} = 40$\label{tab: Salt reconstruction} ]{
   \begin{tabular}{ccccccccccccc}
\toprule
\multirow{3}{*}[-2mm]{  \makecell{number \\ of \\ gliders}} & \multicolumn{12}{c}{Glider Path Design}                                                                                 \\ \cmidrule[.5pt]{2-13} 
                                     & \multicolumn{3}{c}{parallel} & \multicolumn{3}{c}{parallel90} & \multicolumn{3}{c}{center} & \multicolumn{3}{c}{cross} \\
                                     \cmidrule(rl){2-4} \cmidrule(rl){5-7} \cmidrule(rl){8-10} \cmidrule(rl){11-13}
                                     & length   & RMSE    & Corr.   & length   & RMSE     & Corr.    & length  & RMSE    & Corr.  & length  & RMSE   & Corr.  \\ \midrule[.5pt]
4  & 1027 & 0.0568 & 0.9645 & 845  & \textbf{0.0529} & \textbf{0.9692} & 1449 & /               & /               & 936  & 0.0678 & 0.9494 \\
5  & 1283 & 0.0411 & 0.9816 & 1056 & 0.0423          & 0.9803          & 1449 & \textbf{0.0366} & \textbf{0.9857} & 1193 & 0.0574 & 0.9662 \\
6  & 1540 & 0.0338 & 0.9876 & 1268 & 0.0339          & 0.9874          & 1872 & \textbf{0.0279} & \textbf{0.9915} & 1404 & 0.0461 & 0.9771 \\
7  & 1797 & 0.0251 & 0.9933 & 1479 & 0.0257          & 0.9928          & 1872 & \textbf{0.0222} & \textbf{0.9946} & 1661 & 0.0379 & 0.9844 \\
8  & 2053 & 0.0220 & 0.9948 & 1690 & 0.0214          & 0.9951          & 2385 & \textbf{0.0153} & \textbf{0.9975} & 1872 & 0.0310 & 0.9895 \\
9  & 2310 & 0.0189 & 0.9961 & 1901 & 0.0183          & 0.9964          & 2385 & \textbf{0.0146} & \textbf{0.9977} & 2128 & 0.0224 & 0.9945 \\
10 & 2567 & 0.0153 & 0.9975 & 2113 & 0.0150          & 0.9976          & 2808 & \textbf{0.0143} & \textbf{0.9978} & 2340 & 0.0188 & 0.9961 \\ \bottomrule
\end{tabular}}

   \subfloat[Temperature field reconstruction under different number of blocks $B_{\text{dep}}$ and different number of gliders under ``parallel" design with  $B_{\text{long}} = B_{\text{lat}} = 3$\label{tab: Block num Comparison}]{\begin{tabular}{ccccccccc}
\toprule
\multirow{3}{*}[-2mm]{  \makecell{number of gliders}} & \multicolumn{8}{c}{$B_{\text{dep}}$, the number of blocks in depth}                                                             \\ \cmidrule[.5pt]{2-9}  
                         & \multicolumn{2}{c}{10}     & \multicolumn{2}{c}{20} & \multicolumn{2}{c}{40} & \multicolumn{2}{c}{80} \\ 
                         \cmidrule(rl){2-3} \cmidrule(rl){4-5} \cmidrule(rl){6-7} \cmidrule(rl){8-9}
                         & RMSE           & Time      & RMSE     & Time        & RMSE     & Time        & RMSE     & Time        \\ \midrule[.5pt] 
4 & 0.371 & 14.76   min & 0.376 & 3.99   min & 0.383 & 1.33   min & 0.392 & 43.63 sec \\
5 & 0.233 & 26.32 min   & 0.237 & 7.67 min   & 0.244 & 2.26 min   & 0.252 & 1.17 min  \\
6 & 0.184 & 1.44 hour   & 0.190 & 24.41 min  & 0.196 & 7.72 min   & 0.208 & 3.05 min  \\
7 & 0.122 & 1.06 hour   & 0.128 & 19.48 min  & 0.134 & 5.24 min   & 0.144 & 2.21 min  \\
8 & 0.101 & 1.91 hour   & 0.106 & 34.98 min  & 0.112 & 9.30 min   & 0.119 & 3.94 min \\ \bottomrule
\end{tabular}}
   \label{tab:zong}
\end{table}

The  reason for the center design having the lowest reconstruction RMSEs for salinity for all numbers of gliders  while it was not always the best for temperature reconstruction was due to the structure of the underlying 
salinity field was more circular shaped than that of the temperature field  as shown in Fig \ref{fig: TSField}, especially in the shallow depth. The center design had been  commonly employed  in ocean glider surveys.   

Overall speaking the difference between the reconstruction RMSEs among the three formation: center, parallel and parallel90 were not very large with the latter two designs affording smaller effort. From the perspective of practical oceanic implementation, the parallel and parallel90 formations offer easier gliders release and collection logistic than the center formation, as well as require less glider travel effort. {It is noted that for the center design, due to the rectangle shape of the survey area and the equal angle allocation, 
%path lengths vary by direction (with diagonal paths being the longest), resulting in 
the total travel distances with  2\(k\)  gliders is the same with that with 2\(k+1\) gliders. However, the sample coverage with 2\(k+1\) gliders is greater than that of 2\(k\) gliders,  
%as the distance between glider travel paths was reduced, 
leading to a reduction in reconstruction error.}  
%\fn{? (Wu: I want to explain about center design why 4/5, 6/7, 8/9 are equal length but error %decrease)} 

Table \ref{tab:zong} (c) shows that when the number of vertical block $B_\text{dep}$  was increased, the amount of reconstruction time was much reduced with the reduction rate being more than halved  when  $B_\text{dep}$ was doubled each time, while the RMSEs were only slightly increased. The increase in the RMSE was small when compared with the substantial amount of computational saving. {For the parallel design, if the number of gliders was increased from 4 to 8, the reconstruction RMSE of  the temperature field decreased from 0.3831 to 0.1124, which was only $1/3$ of that with 4 gliders.}  

\begin{figure}[p]
  \centering
  \includegraphics[width = 0.47 \textwidth]{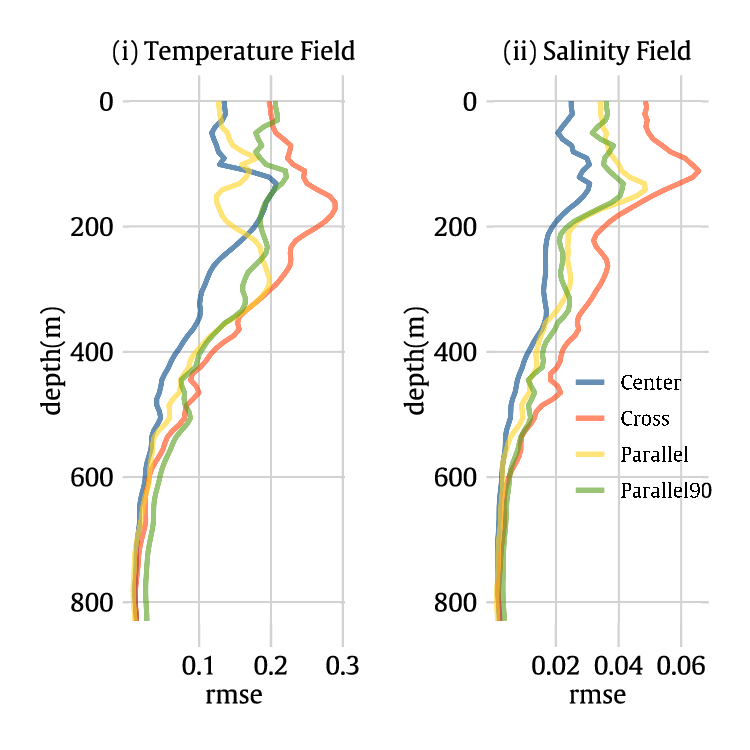}
  \includegraphics[width = 0.47 \textwidth]{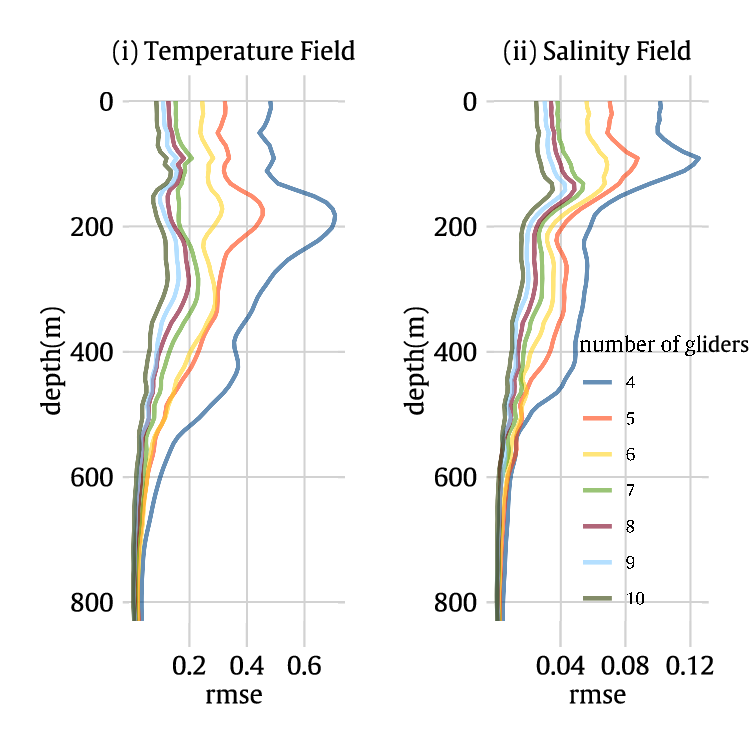}
  \vskip -20pt
  \subfloat[Different glider path designs with 8 gliders \label{fig:RMSEvsPlan}]{\hspace{0.47\linewidth} }
  \subfloat[ Different number of gliders under  ``parallel" design \label{fig: RMSEvsDepth}]{\hspace{0.47\linewidth} }
  
  \caption{Vertical reconstruction RMSES with respect to the depth for temperature ($^\circ \text{C}$, left panels)  and  salinity (g/kg, right panels)  fields with different path designs (a)  and  different number of gliders (b). \label{fig: RMSE}}
\end{figure}

{Figure \ref{fig: RMSE} displays the reconstruction errors of the hydrographic fields at different depths using different path designs with 8 gliders (Panel (a))  and varying numbers of gliders under the parallel design (Panel (b)).  The vertical distribution of the reconstruction error largely reflected 
the vertical gradients of the eddy, which is less variable
as it gets deeper. The largest reconstruction error occurred  between 100 and 300 meters, where the thermocline is located \cite{LI2020101893}. }
\iffalse 
. However, it is not monotone with the largest variation occurs at
a depth around the 150-200 meter range. 
The results show that reconstruction errors were relatively small at depths above 400 meters. However, at depths between 100-300 meters, where the thermocline is located, the temperature gradient is steep, leading to larger reconstruction errors \cite{LI2020101893}.
\fi % \fn{R3Q4, error is quite unevenly distributed with depth}
{The center design achieved better reconstruction results for both temperature and salinity fields. For the parallel design, reconstruction errors decreased uniformly across depth as the number of gliders increased. However, when the number of gliders reached 7, the additional benefit of adding more gliders was not significant.} {We also tested three more active eddies in the Kuroshio Extension region as shown in Figure \ref{supp-fig:three_eddies} in the SM. The detailed results are presented in Section \ref{supp-sec:res-design} of the SM, and the conclusions are consistent with the findings conveyed in Table II and Fig 5. This indicated the robustness and the general applicability of the proposed algorithm}. 
%, and that  % is rob are effective even for more dynamic and active eddies. 
%Given that our study includes relatively large and active eddies, many of the results are likely generalizable to other mesoscale eddies. } 
% \fn{R3Q4, R4Q4. more eddies}

\begin{figure}[p]
  \centering
  \subfloat[Path diagrams for the five missions \label{fig:Eddy0}]{
    \includegraphics[width = 0.48 \textwidth]{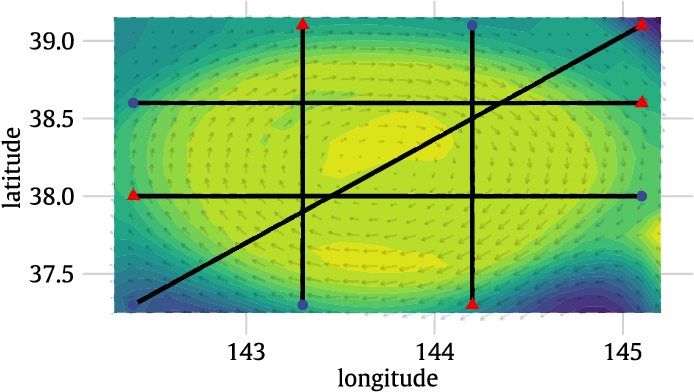}
  }
  \hfill
  \subfloat[Velocity ratios with respect to depth   \label{fig: emp velo field}]{
    \includegraphics[width = 0.48 \textwidth]{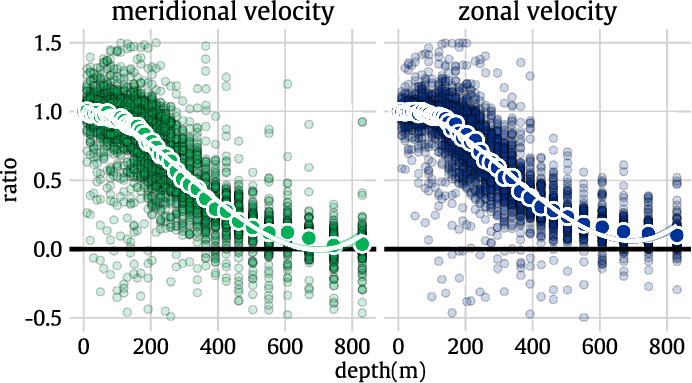}
  }
  \vspace{3mm}
  \caption{(a) Path diagrams for the five missions where the blue dots  and red triangles  represent the start  and end points of the missions, and the arrows represent the ocean current on the sea surface;  and (b) scattered plots of randomly selected 5000 velocity ratios and ocean depth  with the fitted cubic polynomial regression for zonal velocity (left panel) and meridional velocity (right panel) where the average ratios at different depths are marked by bigger circles. 
  The ratios larger than 1.5 or less than -0.5 are excluded (175 for meridional velocity, 74 for zonal velocity). }
\end{figure}

\section{Experimental Results for Path Control} \label{Control Experimental Results}

In this section, we apply Algorithm \ref{alg:alg2} on five different missions and solve the glider path control sub-problem with some evolution algorithms. {We used the same simulated mesoscale eddy and associated hydrographic field from CESM, as in Section \ref{Experimental Results}, but with an expanded time period from March 18 to March 28, 1989, serving as the underlying ground truth.}
% \fn{do we need to mention this again ? (Wu: I omit the region, but the time period is expanded compared to what we used in path design)}  
We experimented five missions whose start and terminal points are reported in Table \ref{tab:tasks} with their pathways displayed in Fig \ref{fig:Eddy0}. {We used the daily average velocity field from CESM in  
%\fn{how to get the filed ? Assimilated data from a product ? (Wu: I have introduced CESM in Glider Path Design, then just say data souce here))}  
the simulation. When the simulation advances to the next day, we immediately switch to the velocity field of the next day to be consistent with %the situation in  
an actual mission.}%\fn{R1Q3, R4Q10. time evolving}

\begin{table}[ht]
\centering
\caption{Coordinates of the start and end points of the line paths  of the five missions}
\label{tab:tasks}
\begin{tabular}{ccc}
\toprule
Mission ID & Start Point            & End Point         \\ \midrule
1       & {(}145.1$^\circ$E, 38.0$^\circ$N{)}   & {(}142.4$^\circ$E, 38.0$^\circ$N{)}   \\
2       & {(}142.4$^\circ$E, 38.6$^\circ$N{)} & {(}145.1$^\circ$E, 38.6$^\circ$N{)} \\
3       & {(}143.3$^\circ$E, 37.3$^\circ$N{)} & {(}143.3$^\circ$E, 39.1$^\circ$N{)} \\
4       & {(}144.2$^\circ$E, 39.1$^\circ$N{)} & {(}144.2$^\circ$E, 37.3$^\circ$N{)} \\
5       & {(}142.4$^\circ$E, 37.3$^\circ$N{)} & {(}145.1$^\circ$E, 39.1$^\circ$N{)} \\ \bottomrule
\end{tabular}
\end{table}

We consider the following algorithms to carry out  the optimalization sub-problem in the path control. 
They were the Differential Evolution (DE) algorithms and the Self-adaptive DE algorithms, called JDE \cite{2006Self}, with three evolution strategies 1-3  (see Part \ref{supp-sec:de} of the SM for details). Among the three evolution strategies, strategy 1 emphasizes the survival of the fittest, strategy 3 focuses on diversity, and strategy 2 is a balance of the strategies 1 and 3.  The Self-Adaptive Multi-population Differential Evolution (SAMDE) algorithm  \cite{2020A}, which is conducive to increasing the diversity of the evolving population, was also considered. In addition, the DEPSO algorithm which combines particle swarm optimization (PSO) with DE was considered too \cite{488968, WANG2019105496}.  Compared with the DE algorithm, the PSO is a non-monotonic search strategy, and combining it with the DE aspect can greatly increase the diversity of the population. The parameter settings for the DE and JDE algorithms are given in Table \ref{supp-tab:de_parameters} of the SM.

\subsection{Ocean Current Field Imputation} \label{subsec: EVF}

Solving the objective function \eqref{eq:weighted average} required real-time ocean current data. However, only the surface ocean current can be obtained by satellite remote sensing \cite{1470035,He2016}. Surface ocean current measurements had been considered as an useful input for the underwater glider control \cite{ALLOTTA201713668}. Several studies showed that there is a correlation between deep-water currents and surface currents \cite{Zhu20, Purkiani20}. 
We used here a statistical approach that imputes the  ocean current velocity with the  surface current measurements as described below. {In practice, near-real-time sea surface velocity can be obtained by calculating geostrophic velocity fields from altimetry satellite observations, such as those provided by Global Ocean Gridded L 4 Sea Surface Heights \cite{DUACS}. 
} % \fn{R1Q4, R4Q2, surface velocity field}

\begin{table}[h]
\centering
\caption{The results of the cubic polynomial regression of the zonal velocity ratio and meridional velocity ratio with parameter estimates, standard errors(std.err), $t$-test results, and the goodness-of-fit of model as well as the $F$-test results. \label{tab: ratio regression}}
\begin{tabular}{c|cccccccc}
\toprule
      direction                     &             & Estimate  & std.err       & $t$-stat  &  $p$-value     & $R^2$                     & $F$-stat                    &  $p$-value                    \\ \midrule
\multirow{4}{*}{zonal} & (Intercept) & 1.03      & 3.08E-03     & 335.51 & 0        & \multirow{4}{*}{0.774} & \multirow{4}{*}{2.64E+04} & \multirow{4}{*}{0} \\
                           & $x_{k,3}$          & -3.84E-04 & 3.80E-05 & -10.11 & 5.42E-24 &                        &                           &                    \\
                           & $x_{k,3}^2$        & -4.93E-06 & 1.20E-07 & -41.19 & 0        &                        &                           &                    \\
                           & $x_{k,3}^3$        & 4.98E-09  & 1.02E-10 & 48.85  & 0        &                        &                           &                    \\ \midrule[.6pt]
\multirow{4}{*}{meridional}  & (Intercept) & 1.03      & 3.85E-03     & 267.00 & 0        & \multirow{4}{*}{0.711} & \multirow{4}{*}{1.84E+04} & \multirow{4}{*}{0} \\
                           & $x_{k,3}$          & -6.27E-04 & 4.75E-05 & -13.21 & 1.05E-39 &                        &                           &                    \\
                           & $x_{k,3}^2$            & -4.62E-06 & 1.50E-07 & -30.79 & 0        &                        &                           &                    \\
                           & $x_{k,3}^3$            & 4.87E-09  & 1.28E-10 & 38.14  & 0        &                        &                           &                    \\ \bottomrule
\end{tabular}
\end{table}

\begin{figure}[h]
  \centering
  \subfloat[Realized paths for Mission 1 (weak current) \label{fig:Path1}]{\hspace{.5\linewidth}}
  \subfloat[Realized paths for Mission 5 (strong current) \label{fig:Path5}]{\hspace{.5\linewidth}}{
    \includegraphics[width = \textwidth]{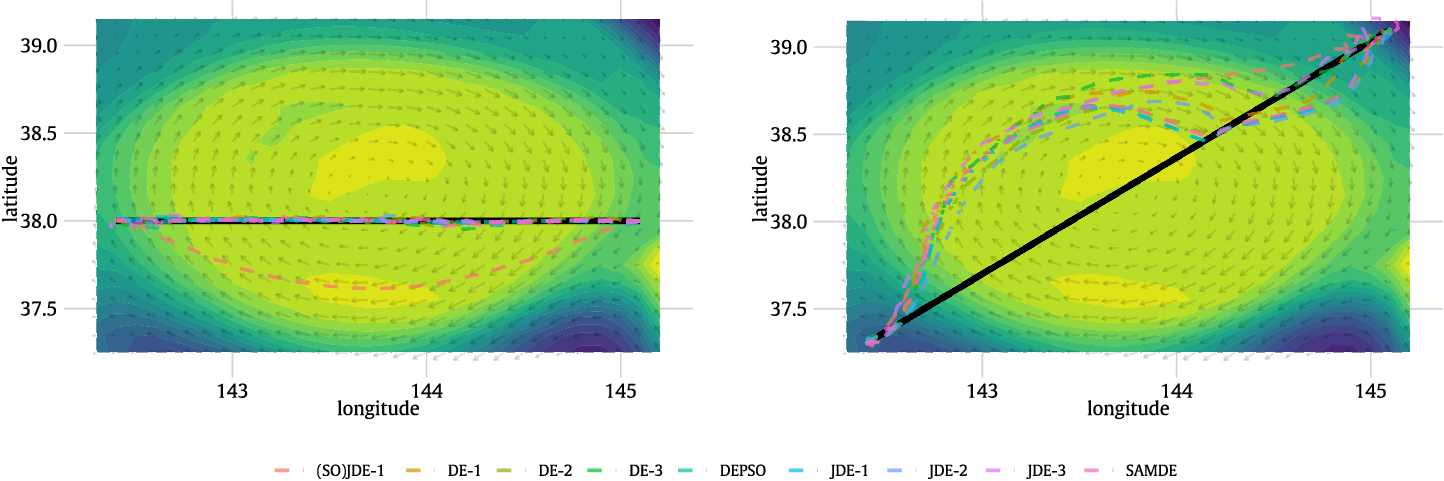}
  }
  
\vspace{2mm}
\caption{The paths generated by solving the glider path control subproblem with different algorithms on mission 1 (a) and mission 5 (b), where (SO) means that the single objective function \eqref{eq:minD1_rewrite} is adopted. \label{fig: glider control paths}}
\end{figure}

% As we have access to the ocean current data\fn{how to get this ? which product ?} of the simulated eddy, which we treat as the underlying truth, we would like to formulate an approach\fn{approach ? } based on the eddy's current observations from February 16 to March 17, 1989, over the study region with $23400$ grids as described earlier. {\bf In this region, the maximum absolute horizontal velocity was 1.28 m/s.} \fn{R3Q5, absolute values of the horizontal velocity} Thus, over the 30 days, we had a total of 702,000 observations, which were used to train a model for estimating the full current velocity field from the surface current data.

{
%Before surveying the eddy, we have access to ocean reanalysis data from recent periods (e.g., near-real-time analysis and forecast data provided by GLORYS), which can often be matched to eddies observed in the real ocean. In this study, for example, 
We used the eddy data from CESM over the period (February 16 and March 17, 1989) as the training data to learn the statistical relationship between surface and subsurface ocean current velocities. The learned relationship will be used in the glider path control to impute real-time subsurface velocities when surface velocities are available. As previously mentioned, near-real-time surface ocean current field is accessible in data products like CESM or GLORYS.}  
%\fn{fails to disclose the needed data information (Wu: I have totally rewritten this %paragraph)}

Across the 30-day period and 23,400 grid points in the study region, we obtained a total of 702,000 observations. Let \( v_{\text{long}}(\boldsymbol{x}_{k}) \), \( v_{\text{lat}}(\boldsymbol{x}_{k}) \), and \( v_{\text{depth}}(\boldsymbol{x}_{k}) \) represent the three velocity components at a location \( \boldsymbol{x}_{k} = (x_{k1}, x_{k2}, x_{k3}) \) for \( k = 1, 2, \dots, 702000 \). Our analysis of the simulated eddy data showed that the vertical velocity component in the target region was on the order of \( 10^{-4} \, \text{m/s} \), negligible compared to both the vertical velocity of the glider (0.17 m/s) and the absolute horizontal eddy velocity (1.28 m/s). % \fn{R3Q5, absolute values of the horizontal velocity} 
Therefore, we considered only the two horizontal velocity components, \( v_{\text{long}} \) and \( v_{\text{lat}} \).

Let $\rho _{\text{long},k}=\frac{v_{\text{long}}\left( \boldsymbol{x}_{k} \right) }{v_{\text{long}}\left( x_{k1},x_{k2},0 \right)}$ and $\rho _{\text{lat},k}=\frac{v_{\text{lat}}\left( \boldsymbol{x}_{k} \right) }{v_{\text{lat}}\left( x_{k1},x_{k2},0 \right)}$ be the velocity ratios at depth $x_{k 3}$ over that at the surface. %for $k = 1,2,\cdots, 70200$. 
 Fig \ref{fig: emp velo field} displays scatter plots of the two ratios and their local averages with respect to the depth $x_{k3}$, which show some level of non-linearity. This motivates 
a cubic polynomial regression model for the velocity ratio and the depth: 
\begin{equation} \label{eq: xLR}
   \rho _{\text{long,}k}=\beta _{\text{long,}0}+\sum_{i=1}^3{\beta _{\text{long}, i}x_{k 3}^{i}}+\varepsilon _k. 
\end{equation} 
The least square method  was used to obtained coefficient's estimates $\{\hat{\beta}_{\text{long} ,i}\}_{i=0}^3$. The details of the cubic regression fitting  based on simulated data are reported in Table \ref{tab: ratio regression}, which suggests all the coefficient estimates were statistically significant with quite high fitting $R^2$.    
The fitted cubic regression ratios were given in Fig \ref{fig: emp velo field}.

Utilizing the estimated cubic polynomial regression, for an arbitrary location  $\boldsymbol{x}_0=\left( x_{01},x_{02},x_{03} \right)$ in the ocean, given the real-time ocean current velocity $v_{\text{long}}\left( x_{01},x_{02},0 \right)$  and $v_{\text{lat}}\left( x_{01},x_{02},0 \right)$ on the sea surface, the ocean current velocity at $\boldsymbol{x}_0$ with depth $x_{03}$ can be imputed as 
\begin{equation} \label{eq: emp formula} 
   \hat{v}_{\text{long}}\left( \boldsymbol{x}_0 \right) =\hat{\rho}_{\text{long}}\left( x_{03} \right) \cdot v_{\text{long}}\left( x_{01},x_{02},0 \right), 
\end{equation}
where  $\hat{\rho}_{\text{long}}\left( x_{03} \right) =\hat{\beta}_{\text{long,}0}+\sum_{r=1}^3{\hat{\beta}_{\text{long,}r}x_{03}^{r}}$. And the $\hat{v}_{\text{lat}}\left( \boldsymbol{x}_0 \right)$ can be obtained similarly. These regression equations, which can be used for imputing the current field beneath the ocean surface, are used later in the glider path control experiments. 

\subsection{Path Control}
 
\begin{figure}[h]
  \centering

   \subfloat[Mission 1]{
    \includegraphics[width = 0.48 \textwidth]{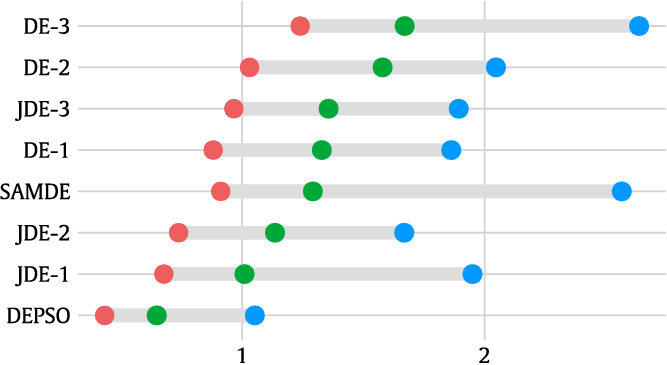}
  }
  \hfill
  \subfloat[Mission 2]{
    \includegraphics[width = 0.48 \textwidth]{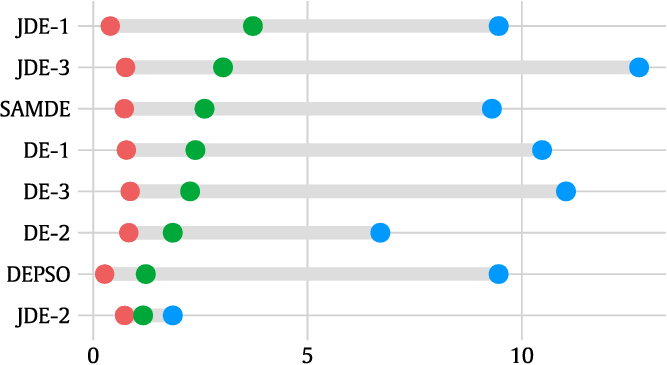}
  }
  
    \vspace{3mm}
  
   \subfloat[Mission 3]{
    \includegraphics[width = 0.48 \textwidth]{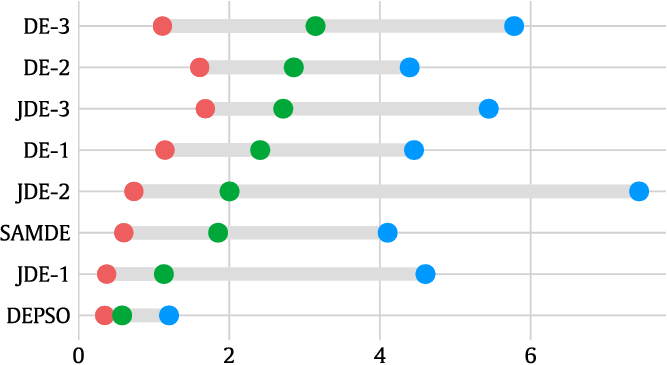}
  }
  \hfill
  \subfloat[Mission 4]{
    \includegraphics[width = 0.48 \textwidth]{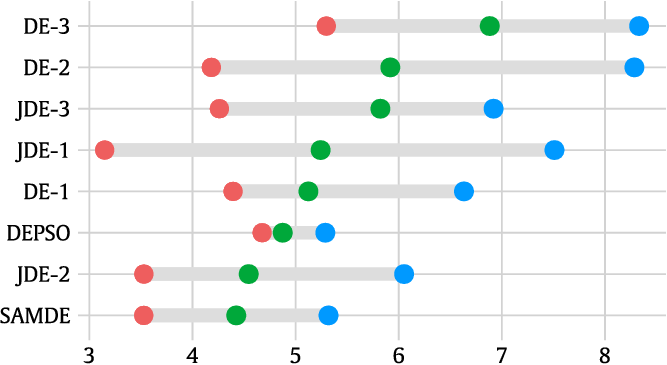}
  }
   \vspace{3mm}
   
     \subfloat[Mission 5]{
    \includegraphics[width = 0.48 \textwidth]{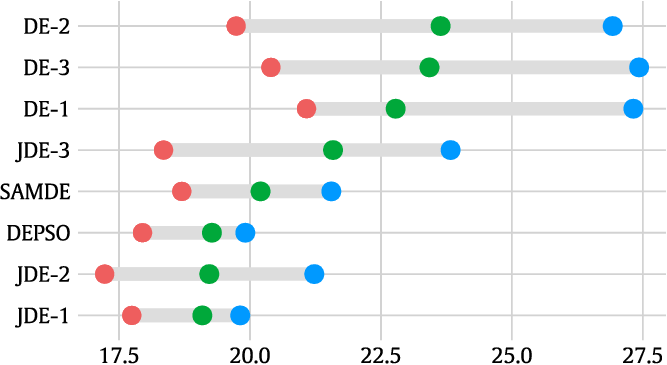}
  }
       \subfloat[Algorithm Rankings \label{fig:sum algo_rank}]{
    \includegraphics[width = 0.48 \textwidth]{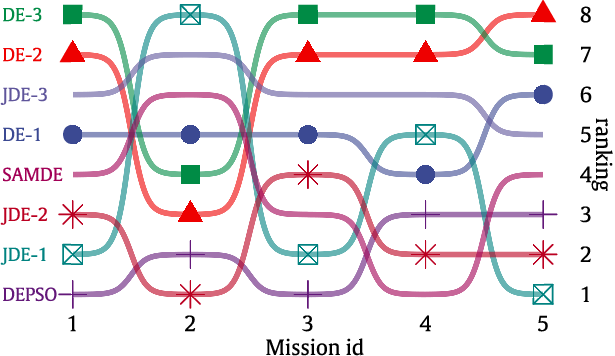}
  }
  
  \vspace{3mm}
  \caption{The minimum, average and maximum deviations (in kms) of the realized paths by the eight algorithms for Missions 1-5 based on 10 replications (Panels (a)-(e)), where red (blue) circles marked the minimums (maximum), and greens were for the averages; And Panel (f) shows  rankings of the algorithms for the 5 missions. \label{fig: deviations}}
\end{figure}

Fig \ref{fig: glider control paths} displays the realized glider pathways administered by the control algorithms for Missions 1 and 5, {which had 
%In our study, all five missions involved  
different travel paths over the same eddy.} {The pathway of Mission 1 was designed in a region with calmer water and weaker currents, while that of Mission 5 traversed areas with stronger currents and greater current shear, making it the most challenging among the five missions. The differences in current strength and shear along these paths contributed to the observed deviation from the prescribed pathways.} % \fn{R3Q6} 
The results for all Missions were reported in Figure \ref{supp-fig:control} in Section \ref{supp-sec:res-control} of the SM. 
The effects of the ocean current was clearly reflected in the realized glider paths and their average deviations from the planned pathway as reported in Fig \ref{fig: glider control paths}, which show  much larger deviations for Mission 5 due to its encountering the strongest current among the five missions. 
The current for Mission 5 was so strong that all the paths directed by the algorithms were ``blown away" from the designated path for a substantial amount of time. However, as the proposed procedure has a built-in component to counter the strong current by increasing weight $w_2$ to the distance to the final destination, the gliders all withstood the strong current and managed to finish the mission without being stuck or  lost all together.  

Fig \ref{fig: deviations} reports the minimum, maximum and average deviations of the eight algorithms based on 10 replications, with the algorithms being ranked from bottom up according to their average deviation. It is observed that the three deviation measures were highly correlated.  Fig \ref{fig:sum algo_rank} reports the ranks of the algorithms based on the average deviations, which shows DEPSO was the best performing algorithm, being the best for two missions and  among the top three for all five missions. JDE-2 was the second best, being the best for one mission and always in the top four, while JDE-1 was the best for the most difficult Mission 5.  It appears that the JDE algorithms were better performing than the DE breed. 
In particular, DEPSO, JDE-1 and JDE-2 occupied the top three spot in terms of the average deviations, while SAMDE ranked the fourth. As reported in Table \ref{supp-tab:control_summary} of SM, 
SAMDE used the least amount of computing time with the average time spent for the optimization per gilder surfacing was 13.2 seconds(s). The average computation time was largely no more than 30s, which was acceptable in the dynamic path control problem. For the best performing DEPSO and JDE-2, the average computing time was 21.2s and 18.5s respectively. More aspects of the results are reported in Section \ref{supp-sec:res-control} of the SM.

\section{Conclusions} \label{Conclusions}

This study considers three key aspects with using the underwater gliders for reconstructing the hydrographic field of a mesoscale eddy.  One is that the TPS interpolation together with the three dimensional blocking scheme offered more accuracy and timely reconstruction than the IDW and Kriging methods. The other is to propose a glider path design method  which reveals that in the absence of ocean current the center and the parallel designs %patterns 
provided better reconstruction of the temperature 
and the salinity fields with 4-10 gliders.  Finally, we put forward a generic glider path control algorithm in the presence of ocean current, which was shown to be able to balance between being approximate to  the pre-designed path when the ocean current is weak and being able to reach the destination when the current is strong.  

\section*{Acknowledgments}

Data were made available through support from NSFC Major Research Plan on West-Pacific Earth System Multispheric Interactions grant 92358303. This research was also partially supported by the National Natural Science Foundation of China grants 12292980 and 12292983.

\bigskip

\bibliographystyle{IEEEtran}
\bibliography{biblio}

\newpage

\section*{Biography Section}

\begin{IEEEbiography}[{\includegraphics[width=1in,height=1.25in,clip,keepaspectratio]{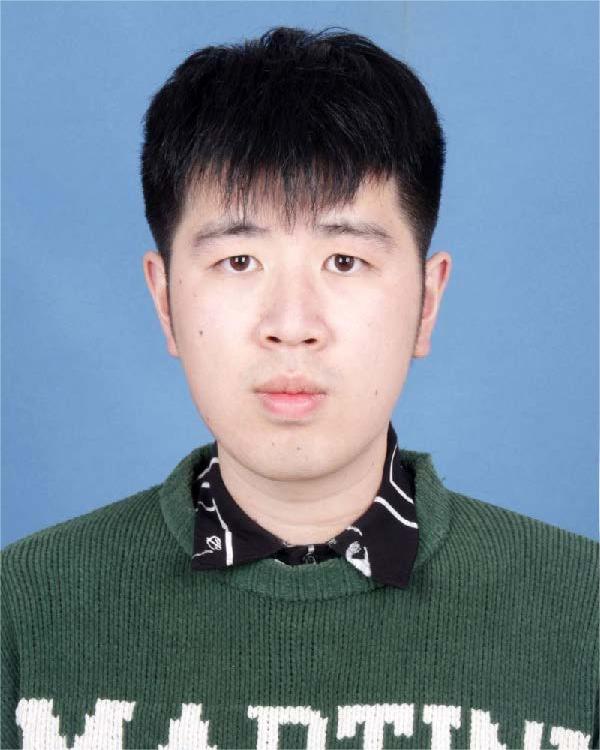}}]{Wu Su}received the B.S. degree from China Jiliang University, Hangzhou,
China, in 2020. He is currently working toward the Ph.D. degree at Peking University, Beijing, China. %  under the supervision of Prof. Song Xi Chen. 

His main area of research is data science. He is interested in the application of artificial intelligence methods to interdisciplinary areas such as oceanic studies and air pollution monitoring. 
\end{IEEEbiography}

\vspace{3pt}

\begin{IEEEbiography}[{\includegraphics[width=1in,height=1.25in,clip,keepaspectratio]{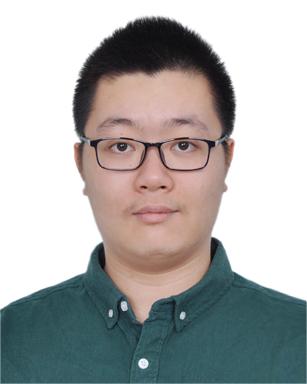}}]{Xiaoyuan E} received his Master of Statistics degree in School for Advanced Interdisciplinary Studies at Peking University, Beijing, China, in 2021. 

He is currently working as an algorithm engineer at JD Logistic. His main area of research is data science and machine learning, and now is in charge of developing algorithm of time series forecasting and anomaly detection in logistic scenarios.
\end{IEEEbiography}

\vspace{3pt}

\begin{IEEEbiography}[{\includegraphics[width=1in,height=1.25in,clip,keepaspectratio]{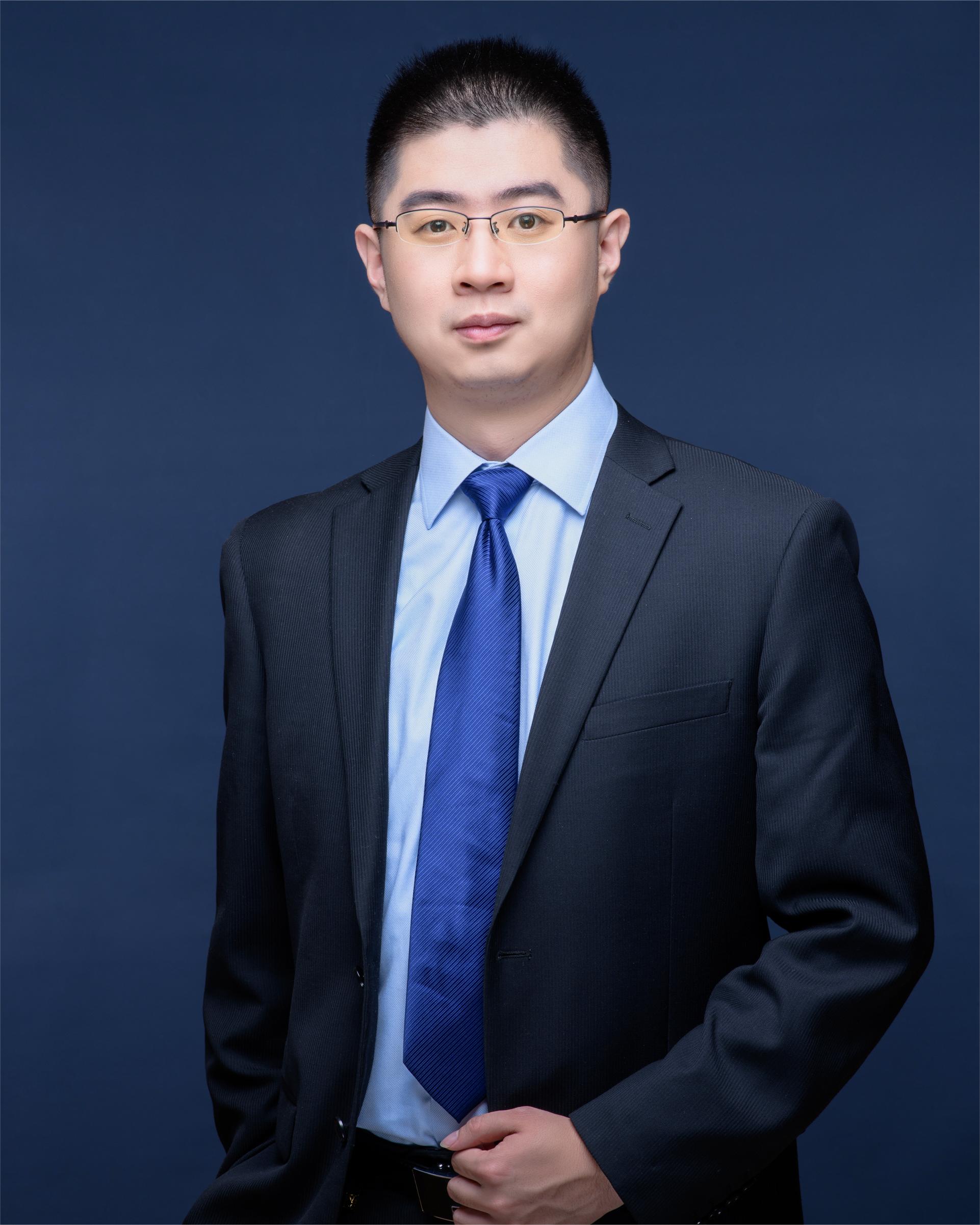}}]{Zhao Jing}received his Ph.D. degree in Department of Oceanography at Texas A\&M University in 2016 and is now a professor at Ocean University of China. 

He has published more than 50 refereed papers. His research interest includes ocean multi-scale dynamical processes and climate impacts. 
\end{IEEEbiography}

\vspace{3pt}

\begin{IEEEbiography}[{\includegraphics[width=1in,height=1.25in,clip,keepaspectratio]{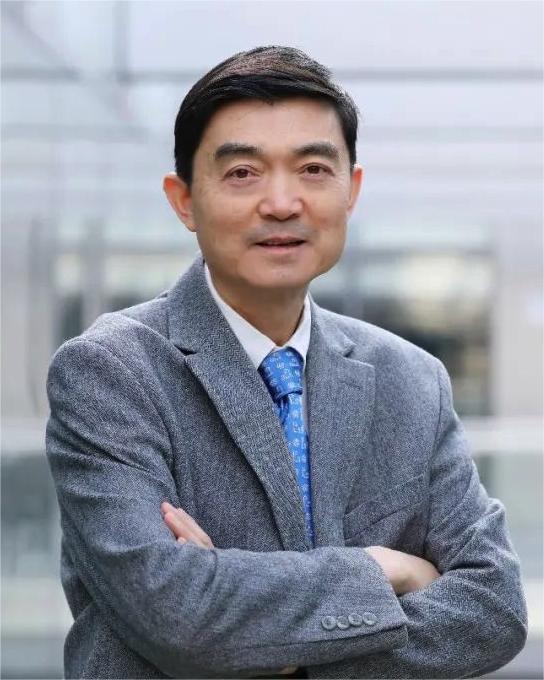}}]{Song Xi Chen}
is a Chair Professor, Peking University. He was elected a fellow of the Institute of Mathematical Statistics (IMS) and American Statistical Association in 2009, a fellow of AAAS in 2018, and a member of Chinese Academy of Science in 2021. He served as a council member of IMS from 2016-2019, and is the scientific secretary of Bernoulli Society for Mathematical Statistics and Probability. He has published more than 110 refereed papers. His research interests include ultra-high dimensional statistical inference, statistical and machine learning, large scale air quality monitoring and assessment and applications in earth science.  
\end{IEEEbiography}

\end{document}